\documentclass[amsmath,amssymb,aps,reprint,superscriptaddress,showpacs,floatfix]{revtex4-2}
\usepackage{epsfig}
\usepackage{color}
\usepackage{graphicx}
\usepackage{amsmath}
\usepackage{mathrsfs}
\usepackage{dcolumn}
\usepackage{bm}

\newcommand{\changed}[1]{\textcolor{black}{#1}}
\begin{document}

\title{Landscapes and nonequilibrium fluctuations of eukaryotic gene regulation}

\author{Masaki Sasai}\email{masakisasai@nagoya-u.jp}
\affiliation{Fukui Institute for Fundamental Chemistry, Kyoto University, Kyoto  606-8103, Japan}
\affiliation{Department of Complex Systems Science, Nagoya University, Nagoya 464-8603, Japan}

\author{Bhaswati Bhattacharyya}
\affiliation{Department of Applied Physics, Nagoya University, Nagoya 464-8603, Japan}

\author{Shin Fujishiro}
\affiliation{Fukui Institute for Fundamental Chemistry, Kyoto University, Kyoto  606-8103, Japan}

\author{Yoshiaki Horiike}
\affiliation{Department of  Applied Physics, Nagoya University, Nagoya 464-8603, Japan}

\date{\today}

\begin{abstract}
Understanding the interplay among processes that occur over different timescales is a challenging issue in the physics of systems regulation. In gene regulation, the timescales for changes in chromatin states can differ from those for changes in the concentration of product protein, raising questions about how to understand their coupled dynamics. In this study, we examine the effects of these different timescales on eukaryotic gene regulation using a stochastic model that describes the landscapes and probability currents of nonequilibrium fluctuations.
This model shows that slow, nonadiabatic transitions of chromatin states significantly impact gene-regulation dynamics. The simulated circular flow of the probability currents indicates a maximum entropy production when the rates of chromatin-state transitions are low in the intensely nonadiabatic regime. In the mildly nonadiabatic regime, this circular flow fosters hysteresis, suggesting that changes in chromatin states precede changes in transcription activity. 
Furthermore, calculations using a model of a circuit involving three core genes in mouse embryonic stem cells illustrate how \changed{the timescale difference} can tune fluctuations in individual genes. These findings highlight the rich effects of nonadiabatic chromatin-state transitions on gene regulation in eukaryotic cells.
\end{abstract}

\maketitle

\section{Introduction}
The coupling of processes with different rates or different timescales is ubiquitous in regulation mechanisms of natural and artificial systems \cite{Nicoletti2024, Honey2007, Hastings2010,  Meykuti2014, Senkowski2024, Ela2012}. Gene regulation is an example of such coupling; the gene activity, or the rate at which the product protein concentration varies, depends on how the chromatin state is modified at different rates. To understand the effects of the rate difference between changes in the protein concentration and the chromatin state, we define the adiabaticity parameter,
\begin{eqnarray}
\omega&=&\frac{\textrm{rate of the chromatin-state change}}{\textrm{rate of the protein-concentration change}}, \nonumber \\
&=&\frac{\textrm{timescale of the protein-concentration change}}{\textrm{timescale of the chromatin-state change}}. \nonumber \\
\end{eqnarray}
Using an analogy from condensed-matter physics, we refer to the system as adiabatic when $\omega > 1$ and nonadiabatic when $\omega< 1$ \cite{Sasai2003, Walczak2005b}. 

\begin{figure}[b]
\centering
\includegraphics[width=9cm]{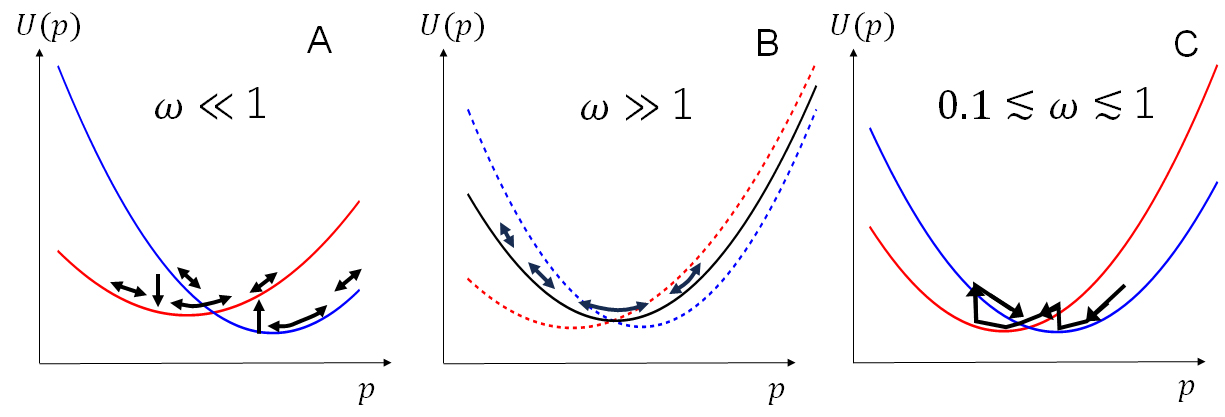}
\caption{
Landscapes explaining the gene expression dynamics. (A) In the nonadiabatic limit of $\omega \ll 1$, the two chromatin states are separately described by the active-state landscape (blue) and the inactive-state landscape (red), and the system dynamics are diffusion over each landscape and the infrequent jumps between two landscapes. (B) In the adiabatic limit of $\omega \gg 1$, the active and inactive states are averaged to give rise to the averaged landscape (black). Dynamics are diffusion over this averaged landscape. (C) In the mildly nonadiabatic case of $0.1\lesssim \omega \lesssim 1$, the frequency of jumps and the diffusion rate are comparable, showing an eddy of probability flow.
}
\label{eddy}
\end{figure}

The difference between adiabatic and nonadiabatic limits can be explained concisely using a landscape picture, which illustrates system dynamics as the movements of the system over landscapes that represent stationary distributions. In the nonadiabatic limit of $\omega \ll 1$, transitions of chromatin between the active and inactive states are infrequent enough to treat the two chromatin states separately. For example, when the chromatin state is kept active (or inactive) with a Poissonian process of protein synthesis, the probability distribution $P(p)$ that cells exhibit the product protein concentration $p$ in each chromatin state should approach a stationary Gaussian-like distribution. The landscape defined by $U(p)=-\log P(p)$ then forms a parabola. 
After the probability distribution becomes stationary, there remain nonequilibrium fluctuations seen in the dynamic system's trajectories over the landscapes. In the nonadiabatic limit, these fluctuations are diffusion over each parabola associated with infrequent jumps between two parabolas  (Fig.\,\ref{eddy}A). In the adiabatic limit of $\omega \gg 1$, on the other hand, the chromatin-state transitions are so frequent that the landscape is effectively averaged between the two states; the fluctuations are diffusion over this averaged landscape  (Fig.\,\ref{eddy}B). It is interesting to examine the mildly nonadiabatic regime of $0.1\lesssim \omega \lesssim 1$, where the rates of transitions and diffusion are comparable, showing the complex pattern of dynamic trajectories (Fig.\,\ref{eddy}C). This complex movement was referred to as ``eddy'' \cite{Walczak2005b}, which was systematically analyzed with a perturbation theory \cite{Shi2011} and an exactly solvable model \cite{Hornos2005}. While the diffusive movements in the cases of $\omega \ll 1$ and $\omega \gg 1$ resemble fluctuations in equilibrium, simplistic analogies to equilibrium are invalid in the regime of $0.1\lesssim \omega \lesssim 1$; therefore, we expect the nonequilibrium features of the gene-switching dynamics to be most evident in this eddy regime.

In Fig.\,\ref{eddy}, we illustrated the coupled dynamics of the chromatin-state change and the protein-concentration change using multiple landscapes corresponding to multiple discrete states of chromatin. By interpreting the transitions between these discrete states of chromatin as changes in continuous variables, and by considering an extended landscape that encompasses both chromatin-state changes and protein-concentration changes, the eddy dynamics can be represented as probability currents over this extended landscape \cite{Zhang2013, Chen2015, Bhaswati2020}. This extended landscape approach provides a physically intuitive description of the eddy dynamics, which we also adopt in the present study.

In studies of bacterial cells, the concept of an adiabatic limit has been widely applied \cite{Tkacik2011}. This adiabatic framework is based on the assumption that the rapid binding/unbinding of transcription factors (TFs) to/from \changed{DNA} ($\approx 10$\,s) \cite{Elf2007} determines the \changed{bacterial chromatin state.} The timescale for changes in protein concentration is comparable to the cell cycle period ($\approx 10^3$\,s) \cite{ZhengH2016},  leading to a ratio of  $\omega\approx 100$, which was thought to validate  the adiabatic limit assumption \cite{Tkacik2011}.
However, several intriguing nonadiabatic phenomena have been observed in bacterial systems. For example, comparing Fig.\,\ref{eddy}A and Fig.\,\ref{eddy}B suggests that the number of basins in the landscape increases as $\omega$ decreases. This trend has been confirmed through experimental manipulations of the binding lifetime of $\lambda$-repressor \cite{Fang2018} and tet-repressor \cite{Jiang2019} \changed{on bacterial DNA.} Moreover, nonadiabatic effects on cell-fate decisions have been emphasized in \textit{Bacillus subtilis} \cite{Schultz2007}.
In \textit{E. coli} cells, single-cell measurements have shown that bursting transitions between active and inactive transcription occur on a timescale of $10^2$ to $10^3$\,s \cite{Golding2005}. These bursting transitions suggest that not only the rapid TF binding/unbinding but also the slower structural changes in bacterial chromatin play a role in chromatin-state transitions. This leads to a mild adiabaticity in the range of $1 \lesssim \omega \lesssim 10$, challenging the conventional assumption of an adiabatic limit. Therefore, it is essential to carefully examine the concepts of adiabaticity and nonadiabaticity in bacterial systems.

Nonadiabaticity is evident in eukaryotic cells, where changes in chromatin states occur due to slow epigenetic modifications of histones.  \changed{In particular, histone modifications such as methylation/demethylation of lysine 9 of histone H3 (H3K9)  or  lysine 27 of  H3 (H3K27) take place over the course of several cell cycles \cite{Hathaway2012, Bintu2016},} with a typical cell cycle period of about $10\sim 20$\,h. In eukaryotic cells, protein concentrations often change due to biochemically regulated active degradation that proceeds on \changed{a timescale of $2\sim 10$\,h \cite{Alber2018, Thomson2011}}, leading to a ratio of  $\omega \approx 0.1$. As a result, eukaryotic genes are typically nonadiabatic \cite{Sasai2013, Ashwin2015, Tiana2016, Fang2018, Folguera2019, Bhaswati2020}. Consequently, in these cells, the eddy dynamics can have a significant impact on gene switching and subsequent cell-fate decisions \cite{Sasai2013, Ashwin2015, Bhaswati2020}. To better understand this phenomenon, a comprehensive analysis using various experimental and theoretical approaches is required.  In this study, we take a step toward this goal by investigating the physical principles underlying the eddy dynamics through a model of eukaryotic gene circuits. We extend the previously developed model for a single eukaryotic gene \cite{Bhaswati2020} to encompass circuits composed of multiple genes. 
\changed{
We apply this model to  the problem of fluctuations in gene switching in mouse embryonic stem (mES) cells. Our findings suggest that the timescale difference in gene regulation can flexibly manage fluctuations in the gene circuit.
}

Eukaryotic chromatin is regulated by various overlapping mechanisms, involving different types of histone modifications and the structural organization of chromatin at multiple spatial scales. \changed{We enhance our model by incorporating multiple degrees of freedom to describe overlapping mechanisms. This extension provides insight into the hypothesis of timescale separation in chromatin regulation and suggests a method for testing the hypothesis.}

\section{A physical model of eukaryotic gene regulation}

\begin{figure}[htbp]
\centering
\includegraphics[width=8cm]{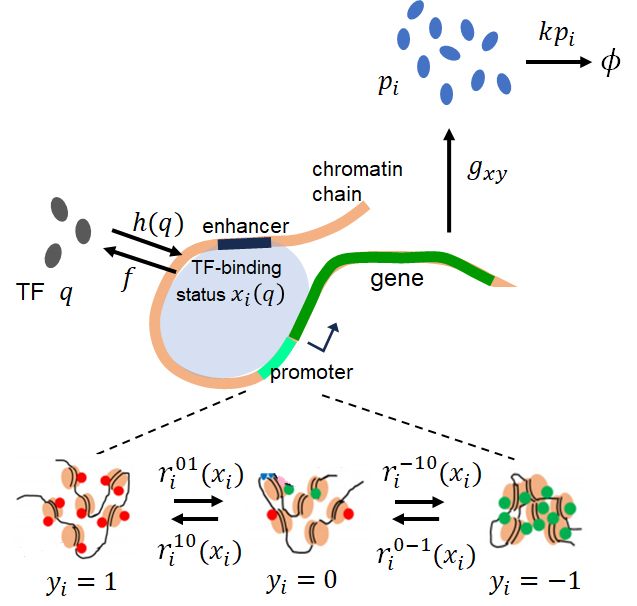}
\caption{
A model of eukaryotic gene regulation. 
The enhancer of an example gene, indexed by $i$, binds a specific transcription factor (TF) with a concentration denoted as $q$ at a rate $h(q)$ and unbinds it at a rate $f$. The likelihood of TFs binding to the gene's enhancer is represented by $x_i(q)$. \changed{The chromatin state $y_i$ represents  modifications of histones} in nucleosomes within the chromatin domain to which the gene's enhancer and promoter belong.
The rate $r_i^{yy'}$ of change from one \changed{chromatin state} $y'_i$ to another state $y_i$ is a function of $x_i$, denoted as $r_i^{yy'}(x_i)$. 
The gene produces a protein with a concentration $p_i$, which degrades with a rate constant $k$.  Processes such as the formation of the transcription-initiation complex (depicted in gray), bursting transcription, translation, and transport are combined into a single process characterized by a rate denoted as $g_{x_i y_i}$. The model can be further extended to include cases where multiple TFs with concentrations $\{q\}=q_1, q_2, \cdots$ bind to the enhancer of the gene.}
\label{scheme}
\end{figure}

In this section, we present a model of eukaryotic gene regulation that emphasizes the differences in timescales between transitions in \changed{states of DNA and chromatin} and changes in protein concentration. The \changed{states of DNA and chromatin of the $i$th gene in the circuit are described using the variables $x_i$ and $y_i$, respectively.} Here, $x_i$ represents the binding status of TFs on the gene's enhancer, while $y_i$ represents the chemical modifications of histones within the chromatin domain that contains the enhancer and promoter of the $i$th gene. The protein concentration is denoted as $p_i$. See Fig. \ref{scheme} for the model scheme. 

We utilize the adiabatic approximation to derive the TF-binding state $x_i$, as explained in the subsection  "Variables and parameters in the model." Therefore, the variables we focus on in our integration of the equations of nonadiabatic dynamics are the histone modification pattern $y_i$ and the protein concentration $p_i$.

\subsection{Variables and parameters in the model}

\subsubsection{Protein synthesis and degradation}
The $i$th gene produces a protein at a rate of $g_{xy}$, resulting in the protein concentration $p_i$. The produced protein is degraded with a timescale of \changed{$1/k= 2\sim10$\,h \cite{Alber2018, Thomson2011}. }
We consider $1/k$ as a typical timescale to change $p_i$.

In the present model, protein production is described as a stochastic process based on the following considerations: (1) In eukaryotic cells, mRNA is transcribed with bursting reactions \cite{Fukaya2023, Porello2023, Meeussen2024}, but the bursting timescale of $10^2\sim 10^3$\,s \cite{Meeussen2024} is shorter than $1/k$. Therefore, we consider transcription to be a continuous stochastic process within our relevant timescale of $1/k$ as a first approximation. 
We will consider transcription bursting explicitly later in the ``Deep epigenetic regulation'' section (Sec. V) in this paper. 
(2)  Processes of mRNA editing, export, and translation, which are involved in protein production, should have shorter timescales than $1/k$. (3) We focus on sufficiently small proteins that can pass through the nuclear pore without significant delay. Based on these points (1), (2), and (3), we adopt a simplified description of protein production as a single-step process occurring at the rate $g_{xy}$. In this notation, the indices $x$ and $y$ represent \changed{the DNA and chromatin states}: $x=1$ (or $x=0$) when the TF binds to (or unbinds from) the enhancer of the gene, and $y=1$ (or $y=-1$) when the histone modification pattern in the chromatin domain containing the gene is active (or inactive). The rate constants used in our model, along with other parameters, are summarized in Table \ref{table:parameters}.

We define the volume of the region containing the chromatin domain of interest as $\Omega$. Given a typical protein concentration $\bar{p}$ as $\bar{p}=g_{11}/(k\Omega)$ when the TF acts as an activator and $\bar{p}=g_{01}/(k\Omega)$ when the TF acts as a repressor, we anticipate that the protein concentration $p_i$ will fall within the range $0 \leq p_i \lesssim \bar{p}$. As the typical protein concentration $\bar{p}$ significantly influences behaviors of gene circuits, we will explore a range of values for $\bar{p}$ in this study.

\begin{table*}
 \begin{center}
   \caption{Parameters in the model of eukaryotic gene regulation}
\begin{tabular}{lccc}  \hline

\multicolumn{2}{c}{Parameters explored in a wide range} & values & references/notes \\ \hline 

Adiabaticity & $\omega$ & $10^{-3}\leq \omega \leq 10^2$  &
\begin{tabular}{c}
$0.1\lesssim \omega\sim r_i^{yy'}/k \lesssim 1$ from the \\ estimation of $r_i^{yy'}$ and $k$.
\end{tabular} 
\\  \\

Typical concentration & $\bar{p}$ & $0<\bar{p}< 10$ & 
\begin{tabular}{c}
$\xi_{11}=\bar{p}$ when the TF is an activator. \\
$\xi_{01}=\bar{p}$ when the TF is a repressor.
\end{tabular}   \\ 
\hline

\multicolumn{2}{c}{Other parameters} & values & references/notes \\ \hline

Rate constant of protein degradation   & $k$ & 1 
& \begin{tabular}{c}
$1/k$ is used as the unit of time. \\
 \changed{$1/k= 2\sim 10$\,h \cite{Alber2018, Thomson2011}} 
\end{tabular}
\\ \\

\begin{tabular}{l}  Rate of protein synthesis \\ 
\changed{at the TF binding state $x$} \\ \changed{and the chromatin state $y$} \end{tabular} 
& $g_{xy}$~~  & $g_{xy}=\xi_{xy}k\Omega$ 
& $\xi_{xy}$ is given in Table II. \\  \\

Volume surrounding chromatin & $\Omega$  & $10^2$ 
&  typical copy number of the protein $\approx \bar{p}\Omega$  \\ \\

\begin{tabular}{l}  Ratio of  \\ 
(binding-rate)/(unbinding-rate) \\ of TF 
\end{tabular}
& $h/f$ & 
\begin{tabular}{c}
$h_0p^2/f$ (\textit{Circuit A}) \\
$h_0p_i^2/f$ (\textit{Circuit B}) \\
$h_0p_1p_2/f$, $h_1p_1p_3^2/f$ (\textit{Circuit C}) \\
\end{tabular}
& 
\begin{tabular}{c}
Values of $h_0/f$ and $h_1/f$ are in Table II, \\ 
which were chosen to make $h\approx f$ when $p_i=\bar{p}$. \\
$f\sim 10^{-1}\text{--}10^0 \,\textrm{s}^{-1}$\cite{Chen2014, Mazzocca2021}
\end{tabular}
  \\ \\
  
\begin{tabular}{l}  \changed{Transition rate of chromatin state }\\from $y'$ to  $y$  around the $i$th gene \end{tabular} & $r_i^{yy'}$ &  \begin{tabular}{c} 
$r_i^{yy'}=\omega_i \bar{r}^{yy'}$ (Eq.\,\ref{scaling})   \\
$\bar{r}^{yy'}=\mu^{yy'}+\gamma^{yy'}x_i$ (Eq.\,\ref{munu})
\end{tabular}
&
\begin{tabular}{c} 
\changed{$r_i^{yy'}\sim 10^{-2}\text{--}10^{-1}\,\textrm{h}^{-1}$ for collective histone }\\
\changed{ methylation/demethylation \cite{Hathaway2012, Bintu2016}.} \\
\changed{$r_i^{yy'}\sim 10^{0}\text{--}10^{1}\,\textrm{h}^{-1}$ for local histone} \\
\changed{acetylation/deacetylation \cite{Bintu2016}.} \\
\changed{$\mu^{yy'}$ and $\gamma^{yy'}$ in the model are given in Table II. }
\end{tabular}\\ 

  \hline
\label{table:parameters}
\end{tabular}
 \end{center}
\end{table*}

\subsubsection{TF binding and unbinding}
The binding status of a specific TF on the enhancer of the $i$th gene is denoted as $x_i$, which plays a crucial role in determining the probability of forming the transcription-initiation complex and the resulting protein-synthesis rate $g_{x_iy_i}$. We write $x_i=1$ when the TF is bound on the enhancer and $x_i=0$ when unbound. Then, $g_{1y}> g_{0y}$ when the TF is an activator and  $g_{1y}< g_{0y}$ when the TF is a repressor. 

We denote the binding rate of the TF to the enhancer as $h(q)$, with $q$ being the TF concentration, and the unbinding rate as $f$. To examine the dynamic effects of the gene on circuit performance, we consider a scenario where $x_i$ is not maintained to 1 or 0. This is facilitated by frequent and alternating binding and unbinding events with $h \approx f$. As a result, the timescale for binding, represented by $1/h$, is similar to the timescale for unbinding, $1/f$. This timescale determines the TF's binding lifetime on chromatin, which has been estimated to be $1/f=1\sim 10$\,s \cite{Chen2014, Mazzocca2021}. Therefore, the timescales of binding and unbinding are significantly shorter than the timescale for changes in protein concentration, \changed{$1/k= 2\sim 10$\,h \cite{Alber2018, Thomson2011}.} The rapid transitions in TF binding and unbinding allow for the use of the adiabatic approximation, leading to the effective equilibration of $x_i$. We use the average value of $x_i$ in this effective equilibrium as $x_i = h/(f + h)$ and regard $x_i$ as a continuous variable of $0\le x_i\le 1$ \cite{Tkacik2011}.

In scenarios where the system includes multiple TFs, we define their concentrations as $\{q\}=q_1,q_2,\cdots$. When the $j$th TF binds to the enhancer of the $i$th gene, we denote the binding rate as $h^{ij}=h^{ij}(q_j)$ and the unbinding rate as $f^{ij}$. The binding state of the $j$th TF on the enhancer of the $i$th gene is represented by $x^{ij}(q_j)$. The adiabatic approximation allows us to express this as $x^{ij}(q_j) = h^{ij}/(f^{ij} + h^{ij}(q_j))$. In this study, we consider that all the activator TFs bound to the enhancer of gene $i$ are essential for expressing that gene, resembling the behavior of an AND gate. Therefore, we define the TF-binding state of the $i$th gene as $x_i = x_i(\{q\}) = \prod_{j} x^{ij}(q_j)$.  Although the model can be applied to other scenarios, such as OR, NOR, and others, we use AND gates in the current study as examples to examine the effects of nonadiabaticity in gene regulation.

\subsubsection{Chemical modifications of histones}
The variable $y_i$ represents whether the chemical modification of histones in a chromatin domain to which the $i$th gene belongs is of the active ($y_i=1$) or inactive ($y_i=-1$) type. As transitions in \changed{the histone modification patterns} can be slower than changes in the protein concentration, we explore the nonadiabatic dynamics of $y_i$ explicitly in the current model.

\changed{In many eukaryotic genomes, sub-megabase (Mb) regions of the chromatin chain are condensed into topologically associating domains (TADs)  \cite{Misteli2020}, which we  refer to as ``chromatin domains''  or simply ``domains.'' These domains serve as functional units and display specific histone modification patterns that reflect the activity of the genes  within them \cite{Misteli2020}. 
}
The chemical modifications of histones in the domain significantly affect the degree of chromatin domain condensation; domains with the inactive histone modifications are more condensed than domains with the active histone modifications \changed{\cite{Bannister2011, Minami2025}.} The degree of domain condensation determines the accessibility of RNA polymerase and related large-sized factors to DNA \cite{Minami2025, Maeshima2015}. 
\changed{Thus, swiching in the histone modification pattern $y_i$ of the domain $i$ leads to a large difference in the protein-synthesis rate $g_{x_i y_i}$ as $g_{x_i -1}\ll g_{x_i 1}$. 
}

\changed{
The kinetics of chemical modifications of histones were quantitatively assessed by recruiting regulatory factors to targeted chromatin regions near specific promoters in mouse and Chinese hamster cells \cite{Hathaway2012, Bintu2016}. These measurements indicated that histone modifications occur at individual nucleosomes within a few hours \cite{Hathaway2012}. Moreover, individual nucleosomes in eukaryotic cells are replaced on a timescale of hours \cite{Deal2010}, which also defines the rate of individual histone modifications.
}

\changed{
The timescale of chromatin-state transitions depends on whether histone acetylation and deacetylation near the promoter determine transcription activity, or if domain-wide patterns of histone methylation and demethylation dictate this activity. In the latter scenario, multiple interacting nucleosomes can collectively modify their methylation patterns through the spreading of methylation within a chromatin domain. This collective modification process takes longer time than modifications at individual nucleosomes \cite{Hathaway2012, Dodd2007, Zhang2014, Chory2019, Sood2020, Owen2023}. 
Regardless of whether local acetylation and deacetylation influence transcription activity or if domain-wide methylation and demethylation play a role, these histone modifications have been shown to lead to all-or-none switching in transcription activity \cite{Hathaway2012, Bintu2016}. In this study, we emphasize this switching as a transition in the variable $y_i$, and we refer to the states represented by $y_i$ as the "chromatin states."
}

\changed{
Here, we} define the parameters $\bar{r}^{yy'}$ of the order of $k$ to characterize transitions of chromatin state $y_i$.  Then, the transition rate $r_i^{yy'}$ from one chromatin state $y'$ to the other $y$ at the $i$th gene is defined as
\begin{eqnarray}
r_i^{yy'}=\omega_i \bar{r}^{yy'}.
\label{scaling}
\end{eqnarray}
\changed{Thus,} $r_i^{yy'}$ is explicitly scaled by the adiabaticity parameter $\omega_i$, showing $r_i^{yy'}=O(\omega_ik)$.

\changed{When the enhancement and disappearance of activating histone acetylation near the promoter play a crucial role, the  measured timescales $1/r_i^{yy'}$ were in hours \cite{Bintu2016}. When the chromatin state is primarily influenced by the spreading and disappearance of repressive histone methylation, these timescales can extend over a few days \cite{Hathaway2012, Bintu2016}. Considering these two scenarios together, the adiabaticity of chromatin state transitions is in the range of $0.1\lesssim \omega_i \lesssim 10$. While such adiabatic ($1\lesssim \omega_i \lesssim 10$) or eddy-regime ($0.1\lesssim \omega_i \lesssim 1$) values of adiabaticity} seem reasonable for eukaryotic cells, we aim to explore a wider range of $\omega_i$ values in this study. This includes both strongly adiabatic ($\omega_i \gg 1$) and strongly nonadiabatic ($\omega_i \ll 1$) cases in order to better understand the dynamics in the eddy regime.

Thus, we examine the circuit dynamics by varying the  parameters $\bar{p}$ and $\omega_i$, while other parameters are fixed to plausible values as experimentally suggested or chosen for simplicity in our simulations as explained in Tables I and II.

\subsubsection{Coupling between the histone and TF-binding states}

The binding and unbinding rates, represented by $h$ and $f$, determine the TF-binding state $x_i$. In the current model, we focus on scenarios where $h$ or $f$ does not explicitly depend on the condensation level of the chromatin domain $y_i$. This lack of dependence is expected when the size of the TFs is sufficiently small, as seen in the case of pioneer factors \cite{Mayran2018}. However, the relationship between $y_i$ and $x_i$ is significant, as transitions in $y_i$ are influenced by $x_i$ as explained below.

The chromatin chain that interacts with activating TFs (activators) recruits histone acetylases. These enzymes alter histone modifications to create an active chromatin state. Conversely, the actions of histone methyltransferases become dominant on the chromatin chain associated with repressive TFs (repressors), which convert the chromatin to an inactive state \cite{Henikoff2011, Ptashne2013}. Thus, the normalized rates $\bar{r}^{yy'}$ and rates $r_i^{yy'}$ of chromatin-state transitions depend on the TF-binding status $x_i$ as $\bar{r}^{yy'}(x_i)$ and $r_i^{yy'}(x_i)$. To simplify this, we apply a first-order approximation regarding the dependence on $x_i$. Using coefficients $\mu^{yy'} \geq 0$ and $\gamma^{yy'}$ of the order of $k$, we  write:
\begin{eqnarray}
\bar{r}^{yy'}(x_i)=\mu^{yy'}+\gamma^{yy'}x_i.  
\label{munu}
\end{eqnarray}
 When the TF is an activator, coefficients for activating transitions are $\gamma^{10} \approx \gamma^{0-1}>0$ and those for inactivating transitions are $\gamma^{-10}\approx \gamma^{01}<0$. When the TF is a repressor, $\gamma^{yy'}$ has the opposite sign as $\gamma^{10}\approx \gamma^{0-1}<0$ and $\gamma^{-10}\approx \gamma^{01}>0$. Values of  $\mu^{yy'}$ and $\gamma^{yy'}$ used in the current study are summarized in Table\,\ref{table:each_circuit}.

Once those $\mu^{yy'}$ and $\gamma^{yy'}$ are given, the adiabaticity parameter $\omega_i$ determines how the chromatin-state transition rate $r_i^{yy'}(x_i)$ depends on the TF-binding state $x_i$ through Eqs.\,\ref{scaling} and \ref{munu}. Since the protein concentration $p_i$ depends on both $x_i$ and $y_i$ via the protein-synthesis rate $g_{x_iy_i}$, Eqs.\,\ref{scaling} and \ref{munu} explain how $p_i$ responds to changes in $x_i$ indirectly through the changes in $y_i$. This indirect response is expected to manifest as a delayed feedback effect, as observed in the heterochromatin bistability in yeast cells \cite{Miangolarra2024}.

\begin{figure}[t]
\centering
\includegraphics[width=8cm]{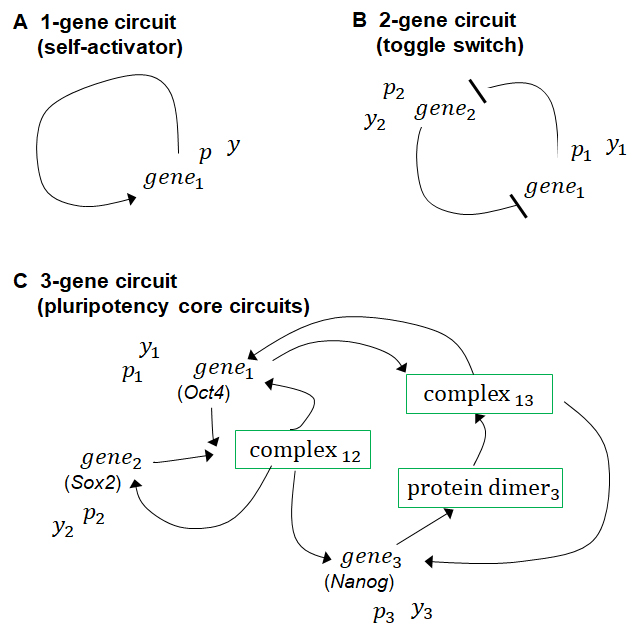}
\caption{
Gene circuits examined in this study: (\textit{Circuit A}) A self-activating single-gene circuit, (\textit{Circuit B}) a mutually repressing two-gene circuit, and (\textit{Circuit C}) a circuit modeling interactions among three core genes, \textit{Oct4}, \textit{Sox2}, and \textit{Nanog}, which maintain pluripotency in mouse embryonic stem (mES) cells. Lines ending with arrowheads represent activating regulation, and lines ending with bars represent repressive regulation.
}
\label{circuits}
\end{figure}

\subsection{System dynamics}
In a previous publication \cite{Bhaswati2020}, we derived the Langevin equations for the dynamics of $p$ and $y$ of a single gene by interpreting both $p$ and $y$ to be continuous variables with $0\le p$ and $-1\le y\le 1$. These equations were derived by representing their master equations in a path-integral form and applying the saddle-point approximation to them. Then,  the landscape of $p$ and $y$, which we call the extended landscape encompassing both the protein-concentration change and the chromatin-state change, was derived to describe their coupled dynamics. Here, in this study, we use this extended landscape method for generic circuits consisting of multiple genes, leading to
\begin{eqnarray}
\frac{1}{k}\frac{dp_i}{dt}&=&G_{p_i}-F_{p_i} +\eta_{p_i},  \nonumber \\
\frac{1}{\omega_i k}\frac{dy_i}{dt}&=&G_{y_i}-F_{y_i}+\eta_{y_i},
\label{chroms}
\end{eqnarray}
where $\omega_i$, which is defined by Eq.\,\ref{scaling}, is the adiabaticity parameter that measures the ratio of the rate to change $y_i$ over the rate to change $p_i$ at the $i$th gene. 
In Eq.\,\ref{chroms}, $G_{p_i}$, $G_{y_i}$, $F_{p_i}$, and $F_{y_i}$ are
\begin{eqnarray}
G_{p_i}&=&\left(\xi_{11}y_{i+}^2+2\xi_{10}y_{i+}y_{i-} + \xi_{1-1}y_{i-}^2\right)x_i  \nonumber \\
&+&\left(\xi_{01}y_{i+}^2+2\xi_{00}y_{i+}y_{i-} + \xi_{0-1}y_{i-}^2\right)(1-x_i),  \nonumber \\
G_{y_i}&=&\left[\bar{r}^{0-1}y_{i-}^2+2\bar{r}^{10}y_{i+}y_{i-}\right]/k, \nonumber \\
F_{p_i}&=&p_i, \nonumber \\
F_{y_i}&=&\left[2\bar{r}^{-10}y_{i+}y_{i-}+\bar{r}^{01}y_{i+}^2\right]/k,
\label{forceC}
\end{eqnarray}
with $y_{i+}=(1+y_i)/2$,  $y_{i-}=(1-y_i)/2$, and $\xi_{xy}=g_{xy}/(k\Omega)$, which is the normalized protein-synthesis rate \changed{at the TF-binding state $x$ and the chromatin state $y$.}

As the saddle-point approximation, i.e., the truncation of fluctuations at the 2nd order, was used in deriving Eq.\,\ref{chroms}, terms $\eta_{p_i}$ and $\eta_{y_i}$ in the r.h.s. of Eq.\,\ref{chroms} represent Gaussian random noises, satisfying $\left<\eta_{p_i}(t)\right>=\left<\eta_{y_i}(t)\right>=\left<\eta_{p_i}(t)\eta_{y_j}(t')\right>=0$,
$\left<\eta_{p_i}(t)\eta_{p_j}(t')\right>=2D_{p_i}\delta_{ij}\delta(t-t')$, and $\left<\eta_{y_i}(t)\eta_{y_j}(t')\right>=2D_{y_i}\delta_{ij}\delta(t-t')$ with $D_{p_i}$ and $D_{y_i}$ being diffusion constants,
\begin{eqnarray}
D_{p_i}&=&\frac{1}{2\Omega}\left(G_{p_i}+F_{p_i}\right), \nonumber \\
D_{y_i}&=&\frac{1}{2\omega_i}\left(G_{y_i}+F_{y_i}\right).
\label{diffusionC}
\end{eqnarray}
From Eq.\,\ref{diffusionC}, we see that the volume $\Omega$ and the adiabaticity $\omega$ determine the fluctuation amplitude of $p$ and $y$, respectively. 

By numerically integrating Eq.\,\ref{chroms},  we calculate the stationary distribution of $\{p\}$ and $\{y\}$, probability currents, and other quantities. In the following, we regard $1/k$ as units of time by adopting $k=1$. See Appendix A for details of the numerical calculation.

We apply this model to circuits of interacting genes.
Biochemical analysis has shown that the slow chromatin-state dynamics bring about prominent feedback effects in eukaryotic cells \cite{Miangolarra2024}. While this experimental analysis is on the gene circuits involving complex repressive regulations, we use in the present study more idealized circuits to highlight the physical principles of the eddy dynamics effects. We examine the \textit{Circuits A, B,} and \textit{C} in Fig.\,\ref{circuits}. Refer to Appendix B and Table II for the details of each circuit.

\changed{In Appendices, we evaluate the characteristics of this model by checking assumptions  used in the model. One assumption examined is on the nonlinearity of the TF binding. As detailed in Appendix B, we highlight the nonlinear effects of TF binding by assuming that each TF in \textit{Circuits A} and \textit{B} is formed as a dimer of the corresponding product protein. In Appendix C, we discuss the landscapes generated by the model under the assumption that each TF is a monomer of the product protein.
The other is the assumption on transitions in chromatin states. The chromatin-state transitions between $y_i = 1$ and $y_i = -1$ occur through an intermediate state, $y_i = 0$ (Fig.\,\ref{scheme}). This intermediate state was emphasized by assuming nonzero parameters for $\xi_{10}$, $\xi_{00}$,  $\bar{r}^{10}$, and $\bar{r}^{-10}$ in Eq.\,\ref{forceC}. In Appendix D, we explain how the landscapes change when we reduce our focus on the intermediate state by modifying Eq.\,\ref{forceC}.}

\section{Landscapes and nonequilibrium fluctuations}
In this section, we investigate the basic features of the effects of nonadiabaticity by applying the model to \textit{Circuits A} and \textit{B}.

\subsection{Basin distribution in nonadiabatic landscapes}

The illustration in Fig.\,\ref{eddy} suggests that the basin distribution of the landscape changes as  $\omega$ varies. We demonstrate that this change indeed takes place in the eddy regime of $0.1\lesssim \omega \lesssim 1$. 

\begin{figure}[b]
\centering
\includegraphics[width=8.1cm]{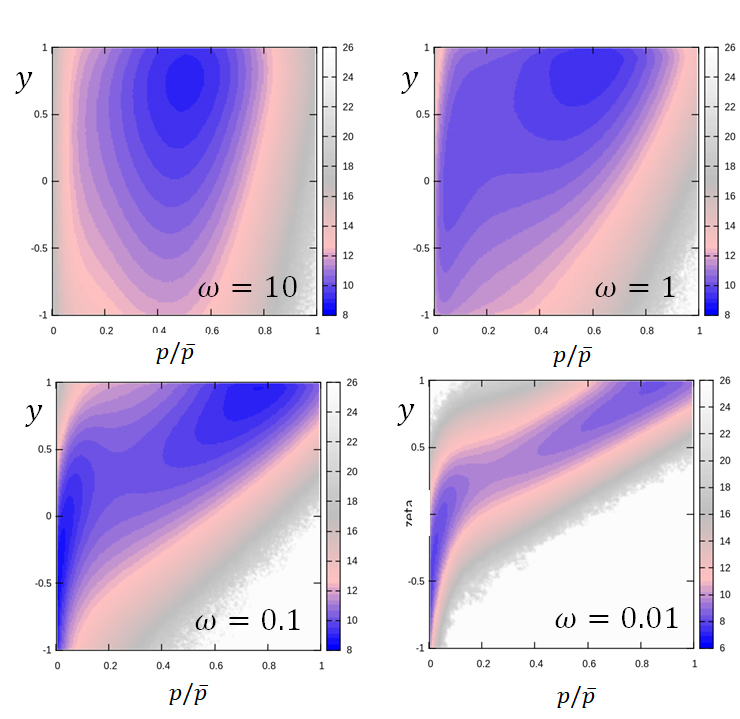}
\caption{
The landscape, $U(p,y)$, of the self-activating single-gene circuit, \textit{Circuit A}, derived from the numerically obtained distribution of $p$ and $y$. $\bar{p}=1.0$.
}
\label{basins1}
\end{figure}

\begin{figure}[htbp]
\centering
\includegraphics[width=8.5cm]{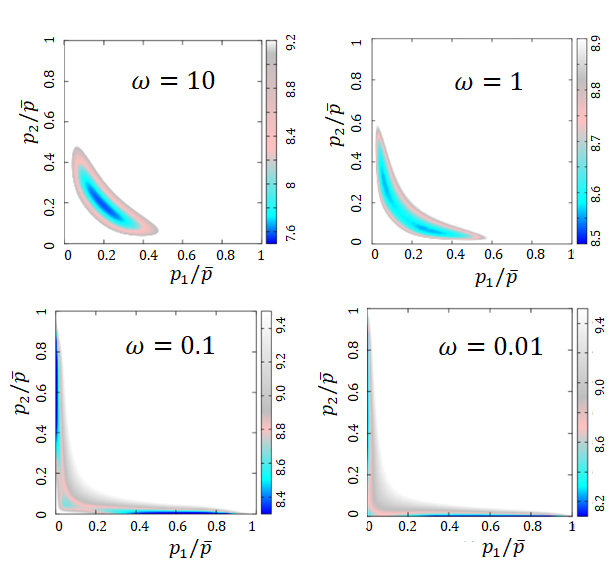}
\caption{
The landscape, $U(p_1,p_2)$, of the mutually repressing two-gene circuit, \textit{Circuit B}, derived by projecting the numerically obtained distribution of $p_1$, $p_2$, $y_1$, and $y_2$ onto the two-dimensional plane of $p_1$ and $p_2$. $\bar{p}=1.2$. Regions of high $U(p_1,p_2)$ values are represented in white. 
}
\label{basins2}
\end{figure}

Fig.\,\ref{basins1} shows the landscape, $U(p,y)=-\log P(p,y)$,  which was obtained from the stationary distribution, $P(p,y)$, in the self-activating single-gene circuit (\textit{Circuit A}). In the large $\omega$ case, the landscape has a single basin at the active state of $y\approx 1$. The basin lies parallel to the $y$-axis, showing a large fluctuation in $y$ but a confined fluctuation in $p$. As $\omega$ decreases, the landscape develops two basins; one at the active state with large $y$ and $p$, and the other at the inactive state with small $y$ and $p$. The inactive basin appears at a distant position from the active basin at the mild nonadiabaticity. In the case of parameters used in Fig.\,\ref{basins1}, the inactive basin appears at  $\omega\approx 0.5$. For the smaller $\omega<0.5$, the landscape extends in both directions of $p$ and $y$, resulting in large correlated fluctuations in $p$ and $y$.

\begin{figure}[htbp]
\centering
\includegraphics[width=9cm]{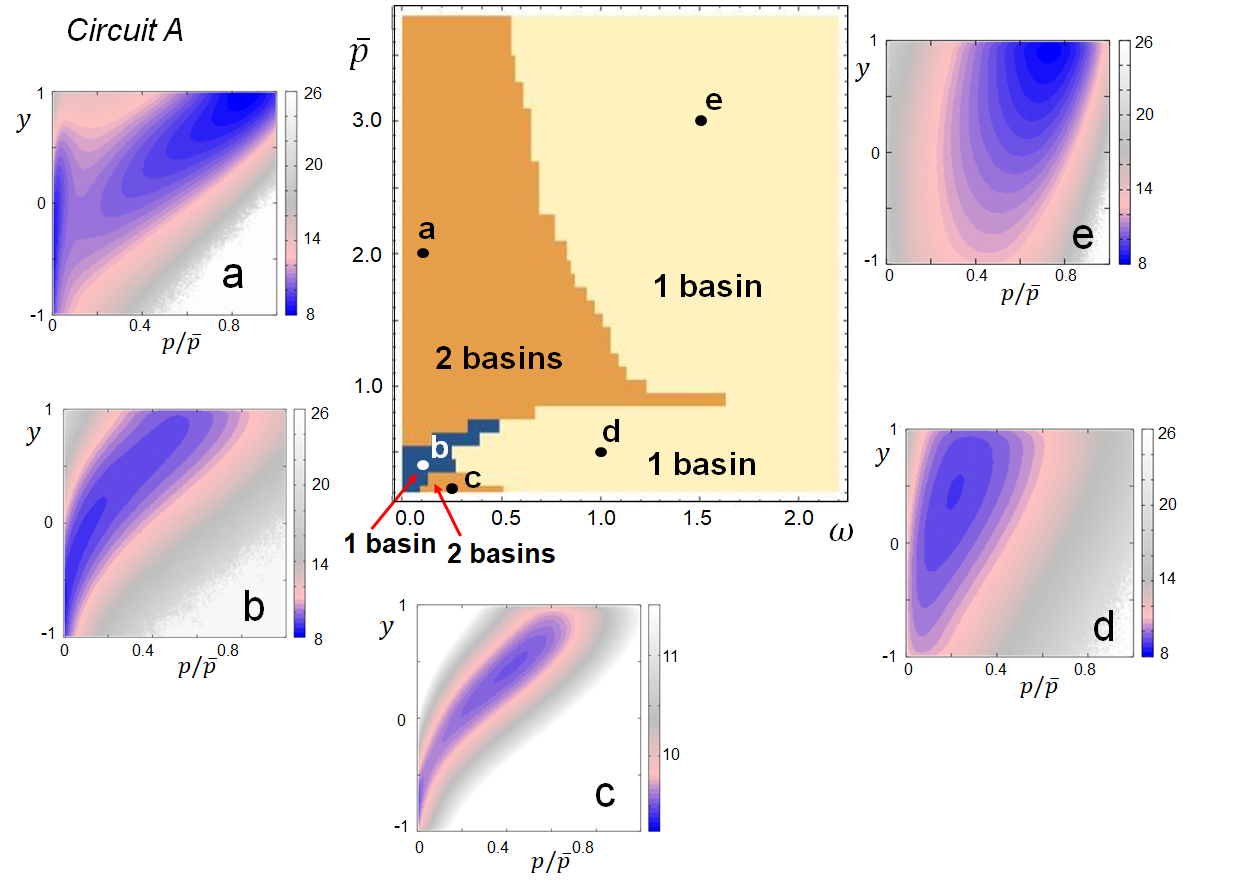}
\caption{Phase diagram of the number of basins on the landscape, $U(p,y)$, of the self-activating single-gene circuit (\textit{Circuit A}).   \changed{Landscapes at representative points are also shown; (a) $\omega=0.1, \bar{p}=2.0$, (b) $\omega=0.1, \bar{p}=0.4$, (c) $\omega=0.25, \bar{p}=0.25$, (d) $\omega=1.0, \bar{p}=0.5$, and (e) $\omega=1.5, \bar{p}=3.0$. } Phase diagrams were drawn by sampling about $10^3$ points on the $\omega\textrm{-}\bar{p}$ plane and counting the number of basins on the numerically obtained steady-state distributions. Lines representing the phase boundaries are not smooth due to the finite number of sampled points.
}
\label{phaseA1}
\end{figure}

\begin{figure}[htbp]
\centering
\includegraphics[width=9cm]{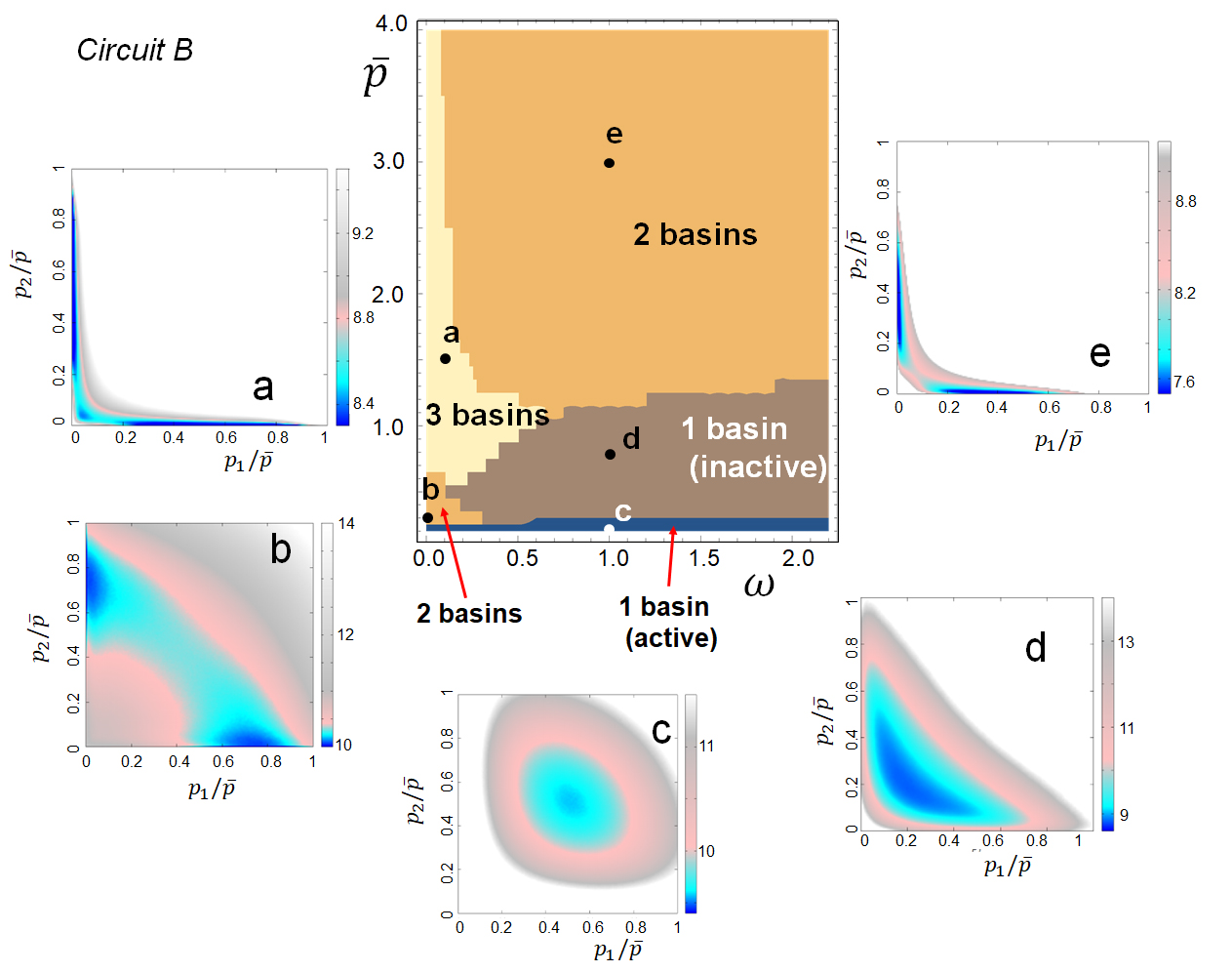}
\caption{Phase diagram of the number of basins on the landscape, $U(p_1,p_2)$, of the mutually repressing two-gene circuit (\textit{Circuit B}).  \changed{Landscapes at representative points are also shown; (a) $\omega=0.1, \bar{p}=1.5$, (b) $\omega=0.01, \bar{p}=0.3$, (c) $\omega=1.0, \bar{p}=0.22$, (d) $\omega=1.0, \bar{p}=0.8$, and (e) $\omega=1.0, \bar{p}=3.0$. } Phase diagrams were drawn by sampling about $10^3$ points on the $\omega\textrm{-}\bar{p}$ plane and counting the number of basins on the numerically obtained steady-state distributions. Lines representing the phase boundaries are not smooth due to the finite number of sampled points.
}
\label{phaseB1}
\end{figure}

Similarly, in the landscape of a mutually repressing two-gene circuit (\textit{Circuit B}), the number of basins increases as the adiabaticity parameter $\omega$ decreases. We numerically calculated the steady-state distribution $P(p_1,p_2,y_1,y_2)$, and plotted the two-dimensional landscape $U(p_1,p_2)=-\log \left( \iint dy_1dy_2P(p_1,p_2,y_1,y_2) \right)$ (Fig.\,\ref{basins2}).
When the typical protein concentration $\bar{p}$ is moderately small ($\bar{p}<1.5$) and the adiabaticity $\omega$ is large, the landscape $U(p_1,p_2)$ has a single basin at the diagonal position, $p_1=p_2$, on the $p_1$-$p_2$ plane, corresponding to the inactive state with small $p$ and $y$. As $\omega$ decreases, the basin at the diagonal position bifurcates into two basins at off-diagonal positions, one at $p_1 > p_2$ and $y_1>y_2$ and the other at $p_1< p_2$ and $y_1<y_2$. With further decreases in $\omega$, the two off-diagonal basins deepen, and an additional shallow basin emerges at the diagonal position, representing the inactive state. In this scenario, the two off-diagonal basins dominate the dynamics, causing the system to function as a toggle switch between those two states. When $\bar{p}$ is large ($\bar{p}>1.5$) in \textit{Circuit B}, the landscape has two basins at off-diagonal positions even with a large $\omega$,  and reducing $\omega$ generates an extra third basin at the diagonal position of the inactive state. Overall, in both scenarios, the landscape extends over multiple basins when $\omega < 1$, indicating increased fluctuations in $p_1$ and $p_2$.

We summarize the characteristics of basins of \textit{Circuits A} and \textit{B} in  phase diagrams \changed{shown in  Figs.\,\ref{phaseA1} and \ref{phaseB1}, respectively. Each figure also illustrates representative landscapes for its corresponding phases.  In both Figs.\,\ref{phaseA1} and \ref{phaseB1},} despite the presence of complex phase-diagram structures at $\bar{p}<1$ due to significant fluctuations in $p_i$, we notice a consistent trend of increasing basin numbers as the parameter $\omega$ decreases, regardless of the $\bar{p}$ value. This rise in basin numbers is observed within the $\omega$ range of the eddy regime of $0.1\lesssim \omega \lesssim 1$. With the increase in basin numbers, the landscape extends across multiple basins, indicating enhanced fluctuations in the expression level.

\changed{
In our current simulations, we have assumed that TFs in circuits exist as dimers of the corresponding proteins. However, if we consider TFs to be monomers instead, the Hill coefficient in the adiabatic expression of $x^{ij}(q_j)$ decreases from 2 to 1. This reduction in nonlinearity diminishes the likelihood of generating multiple basins on the landscape compared to the dimer TF cases. Nevertheless, even in the monomer TF scenarios, the lowering of adiabaticity still broadens the distribution, thereby promoting the tendency to generate multiple basins, as shown in Appendix C. 
}

\subsection{Hysteresis and time-ordering in the eddy regime}

\begin{figure}[t]
\centering
\includegraphics[width=9cm]{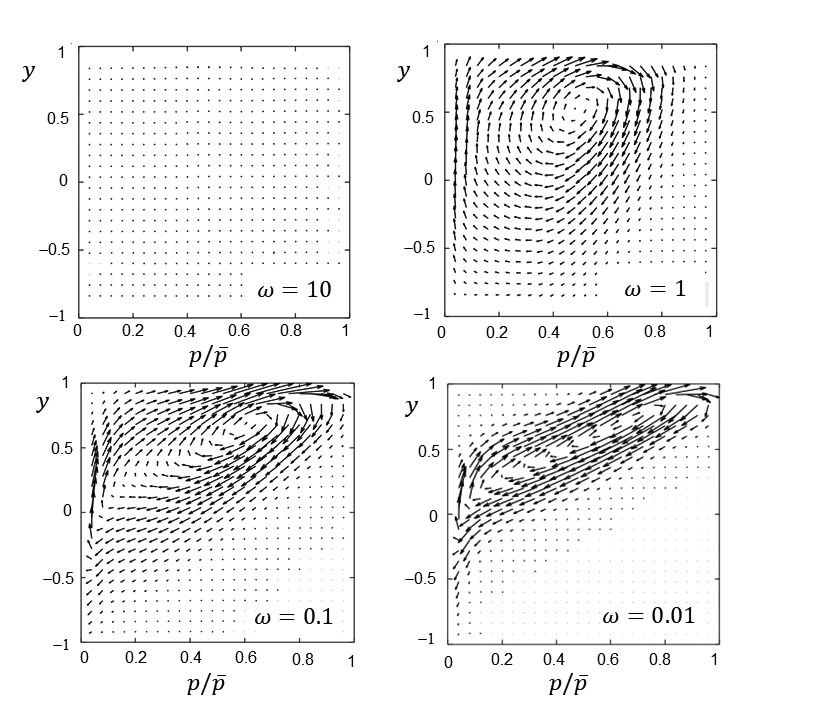}
\caption{
The probability current $\vec{J}(p,y)$ of the self-activating single-gene circuit, \textit{Circuit A}, is plotted as arrows on the $p$-$y$ plane. The same parameters used in Fig.\,\ref{basins1} were employed.
}
\label{singleFlux}
\end{figure}

The changes in the landscape in the eddy regime should alter fluctuations in $p$ and $y$. We can analyze this effect by calculating the probability current, $\vec{J}=(\{J_{p_i}\}, \{J_{y_i}\})$, with

\begin{eqnarray}
J_{p_i}&=&\left( G_{p_i}-F_{p_i}\right)P(\{p\},\{y\}) -\frac{\partial}{\partial p_i}\left[ D_{p_i}P(\{p\},\{y\}) \right], \nonumber \\
J_{y_i}&=&\left( G_{y_i}-F_{y_i}\right)P(\{p\},\{y\}) -\frac{\partial}{\partial y_i}\left[ D_{y_i}P(\{p\},\{y\}) \right]. \nonumber \\
\label{eq:curr}
\end{eqnarray}
The divergence-free circular flow of  $\vec{J}$ in a steady state is a hallmark of broken detailed balance, indicating nonequilibrium dissipation that drives the fluctuation \cite{Wang2008, Feng2011b, Fang2019}.
Fig.\,\ref{singleFlux} shows the probability current in the self-activating single-gene circuit (\textit{Circuit A}).  When $\omega\gg 1$, the current is not very noticeable in the steady state, suggesting that the state can be effectively approximated by an equilibrium. Conversely, in the nonadiabatic case of $\omega\lesssim 1$, the circular flux of the probability current becomes evident,  explicitly breaking the detailed balance. The circular flux spreads globally over the $p$-$y$ plane in the eddy regime of $0.1\lesssim \omega\lesssim 1$, but the area showing the intense current becomes narrower in the strongly nonadiabatic case of $\omega\lesssim 0.01$.

\begin{figure*}[htbp]
\centering
\includegraphics[width=16cm]{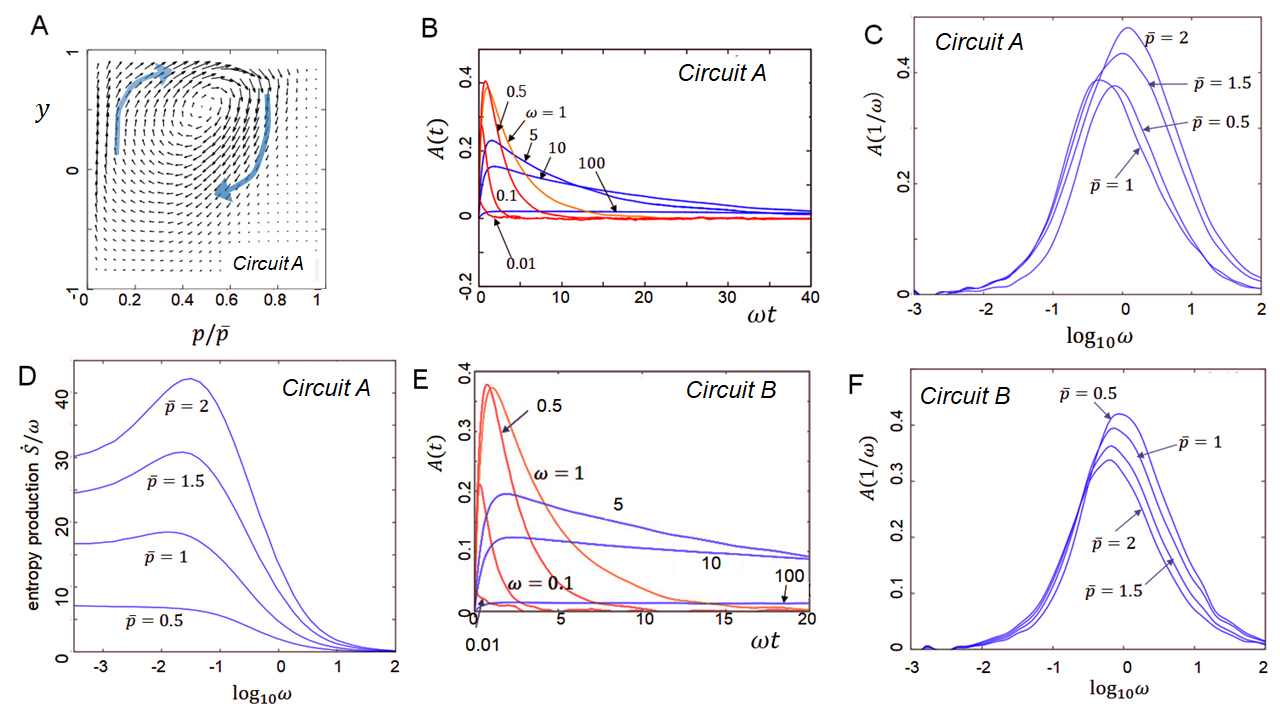}
\caption{
Prominent nonequilibrium features in the eddy regime of $0.1\lesssim \omega \lesssim 1$. (A) The circular flux of the probability current leads to hysteresis in the switching dynamics. The same plot as a panel for $\omega=1$ in Fig.\,\ref{singleFlux}. (B) The cross-correlation function $A(t)$ in the self-activating single-gene circuit, \textit{Circuit A}, is plotted as a function of $t$ for various values of $\omega$; $\omega=1$ (orange), $\omega=5, 10, 100$ (blue), and $\omega=0.5, 0.1, 0.01$ (red). (C) The value of the cross-correlation  $A(t=1/\omega)$  in \textit{Circuit A} is plotted as a function of $\omega$ for various values of $\bar{p}$. (D) Entropy production during the typical chromatin-state turnover time, $\dot{S}/\omega$, in \textit{Circuit A} is plotted as a function of $\omega$ for various values of $\bar{p}$. (E) The cross-correlation function $A(t)$ in the mutually repressing two-gene circuit, \textit{Circuit B}, is plotted as a function of $t$ for various values of $\omega$; $\omega=1$ (orange), $\omega=5, 10, 100$ (blue), and $\omega=0.5, 0.1, 0.01$ (red). (C) The value of the cross-correlation function $A(t=1/\omega)$ in \textit{Circuit B} is plotted as a function of $\omega$ for various values of $\bar{p}$. 
}
\label{hyster1}
\end{figure*}

In the previous publication \cite{Bhaswati2020}, the authors demonstrated that the most probable pathways---those that contribute the most weight in the path-integral representation of the transition probability between the active and inactive states of a self-activating single-gene circuit---align with the direction of the circular flux in its extended landscape. Consequently, the pathway generated in one direction differed from the pathway in the reverse direction, indicating the presence of hysteresis. 

Similarly, when following the direction of the globally developed circular flux in Fig.\,\ref{singleFlux}, we observe that the transition from the inactive state to the active state and the reverse transition should follow distinct pathways.
Therefore, we expect that the global circular flux within the range of \(0.1 \lesssim \omega \lesssim 1\) (Fig. \ref{singleFlux} and Fig. \ref{hyster1}A) in the present system indicates that the transition between the two states exhibits hysteresis. This expectation is  confirmed through the calculation of the cross-correlation function.
 
\begin{eqnarray}
A(t)=\frac{\left<\left( \delta y(\tau)\delta p(\tau +t)-\delta y(\tau +t)\delta p(\tau) \right)\right>_{\tau}}{\Delta y \Delta p}, 
\label{cross}
\end{eqnarray}
where $\left<\cdots\right>_{\tau}$ represents the average over $\tau$, $\delta y(t)=y(t)-\left<y(t_0)\right>_{t_0}$, and $\delta p(t)=p(t)-\left<p(t_0)\right>_{t_0}$. $\Delta y$ and $\Delta p$ are defined by $\Delta y=\left<\delta y(\tau)^2\right>_{\tau}^{1/2}$ and $\Delta p=\left<\delta p(\tau)^2\right>_{\tau}^{1/2}$. If there is a tendency for time-ordered changes with variation in $y$ preceding the  $p$ variation, then $A(t)>0$ for $t>0$. The calculated results in Fig.\,\ref{hyster1}B indeed show $A(t)>0$ for $t>0$, confirming the tendency that the chromatine state $y$ changes first and the product-protein concentration $p$ follows in both transitions from inactive to active and from active to inactive states, resulting in different pathways with hysteresis. $A(t)$ shows a peak at the typical turnover time of the chromatin-state change, $t\approx 1/\omega$. 

In Fig.\,\ref{hyster1}C, $A(1/\omega)$ is plotted as a function of $\omega$, showing that the tendency for cross-correlation, or time-ordering, is most pronounced in the eddy regime of $0.1\lesssim\omega \lesssim 1$. In contrast, this tendency weakens in the strongly adiabatic regime ($\omega \gg 1$). This observation aligns with the results presented in Fig.\,\ref{singleFlux}, where the circular flux develops significantly in the eddy regime but diminishes in the strongly adiabatic regime. In the strongly nonadiabatic regime ($\omega \ll 1$), the cross-correlation $A(1/\omega)$ is also weaker because the probability current becomes confined to a localized area, although the intensity of the current remains high within that area. The narrower distribution of current shown in Fig.\,\ref{singleFlux} for $\omega = 0.01$ suggests that changes in protein concentration tend to closely follow changes in chromatin states. This leads to both active-to-inactive and inactive-to-active pathways approaching a line that connects two basins. The convergence of these pathways should reduce hysteresis, resulting in a smaller value of $A(1/\omega)$.

The amount of nonequilibrium dissipation can be measured by calculating the entropy production rate,

\begin{eqnarray}
\dot{S}=\iint dpdy \left( \frac{J_p(p,y)^2}{P(p,y)D_p}+\frac{J_y(p,y)^2}{P(p,y)D_{y}} \right).
\label{EPR}
\end{eqnarray}
In Fig.\,\ref{hyster1}D, entropy $\dot{S}/\omega$ produced during the chromatin-state turn-over time $1/\omega$ is plotted. Fig.\,\ref{hyster1}D shows that entropy production $\dot{S}/\omega$ remains low in the adiabatic regime of $\omega > 1$, approaching 0 in the adiabatic limit as $\omega \gg 1$. This observation is in line with the finding that the circular flux disappears in the adiabatic limit. 
As $\omega$ becomes smaller, the nonequilibrium dissipation indicated by $\dot{S}/\omega$ begins to rise in the eddy regime of the range $0.1 \lesssim \omega \lesssim 1$. This rise in nonequilibrium dissipation corresponds to a shift in landscape structure, where the circular flux leads to time-ordering tendencies and hysteresis. 

While the cross-correlation $A(1/\omega)$ shows a peak in the eddy regime of $0.1 \lesssim \omega \lesssim 1$, entropy production $\dot{S}/\omega$ continues to increase in this regime as $\omega$ decreases further. This rise is attributed to the increasing intensity of the probability current in the nonadiabatic regime. 
While $A(t)$ reflects the global pattern of circular flux and diminishes as the current becomes confined to a narrow area (as in the case of $\omega=0.01$ in Fig.\,\ref{singleFlux}), entropy production $\dot{S}/\omega$ reflects the intensity of that current and increases as the intensity grows with further decreases in $\omega$. Interestingly, $\dot{S}/\omega$ exhibits a peak in the intensely nonadiabatic regime of $0.01 \lesssim \omega \lesssim 0.1$ and tends to decrease as $\omega$ further decreases, suggesting that the current confined to a narrow area (narrower than shown in Fig.\,\ref{singleFlux}) begins to contribute less to the overall dissipation.

The time-ordering tendency and hysteresis are also observed in the mutually repressing two-gene circuit (\textit{Circuit B}). The cross-correlation, $A(t)$ of Eq.\,\ref{cross}, is plotted in Fig.\,\ref{hyster1}E using $\delta y(t)=y_1(t)-y_2(t)$, $\delta p(t)=p_1(t)-p_2(t)$, showing $A(t)>0$ for $t>0$. $A(1/\omega)$ is plotted in Fig.\,\ref{hyster1}F, demonstrating that the cross-correlation is most prominent in the eddy regime.

Taken together, the examples of single-gene (\textit{Circuit A}) and two-gene (\textit{Circuit B}) circuits illustrate that the landscape exhibits different structures in the adiabatic and nonadiabatic limits with the increased number of basins in the nonadiabatic limit. In the transitional or eddy regime between these limits, fluctuations grow as the probability current shows a well-developed circular flow, which is associated with significant entropy production. This established circular flow results in hysteresis in gene-switching dynamics, leading to a temporal ordering of processes, where the slow change in chromatin state occurs before the fast change in transcription activity.

\section{Fluctuations in the three-gene model}

\subsection{The large Nanog fluctuation in mES cells}
The three core genes, \textit{Oct4}, \textit{Sox2}, and \textit{Nanog}, activate each other and regulate many other genes in mES cells, playing essential roles in maintaining pluripotency \cite{Wang2006, Boyer2006, Niwa2007}. Here, we write the gene's name in \textit{italic} and the protein's name in roman. These genes remain active when cells are cultivated in a medium containing 2i factors \cite{Navarro2018}.  Without 2i, cells still maintain pluripotency when cultivated with Lif, but they show a significant fluctuation in Nanog concentration from cell to cell \cite{Singh2007, Kalmar2009, Canham2010}. These cells show transitions between high and low-Nanog cell states during several cell cycles \cite{Kalmar2009}. As loosing Nanog is a trigger to differentiation \cite{Hyslop2005}, the large Nanog fluctuation is the fluctuation of cells at the doorway to differentiation. However, in the same population of cells, Oct4 and Sox2 do not fluctuate intensely and instead show narrow single-peak distributions of their concentrations \cite{Kalmar2009}.

Models of the gene network were proposed to explain the difference in activity fluctuation among the three core genes \cite{Kalmar2009, Sasai2013, Yu2018, Samanta2019}. A typical assumption used in the models was the self-activating interaction of \textit{Nanog} \cite{Yu2018} and the repressing interactions between \textit{Nanog} and other genes \cite{Kalmar2009, Yu2018}. These assumptions allowed the models to explain the switching fluctuation between active and inactive states in \textit{Nanog} and the lack of relevance of other genes to this fluctuation. However,  the self-repression of \textit{Nanog} was reported \cite{Navarro2012} and the mutual repression among these genes seem to play a minor role \cite{Niwa2007}; therefore, further careful consideration on the gene-circuit structure is necessary to analyze the heterogeneous fluctuations in these genes. In this study, we use our model of chromatin-state transitions to analyze this problem without assuming the mutual repression among these three genes.

\begin{figure}[htbp]
\centering
\includegraphics[width=8cm]{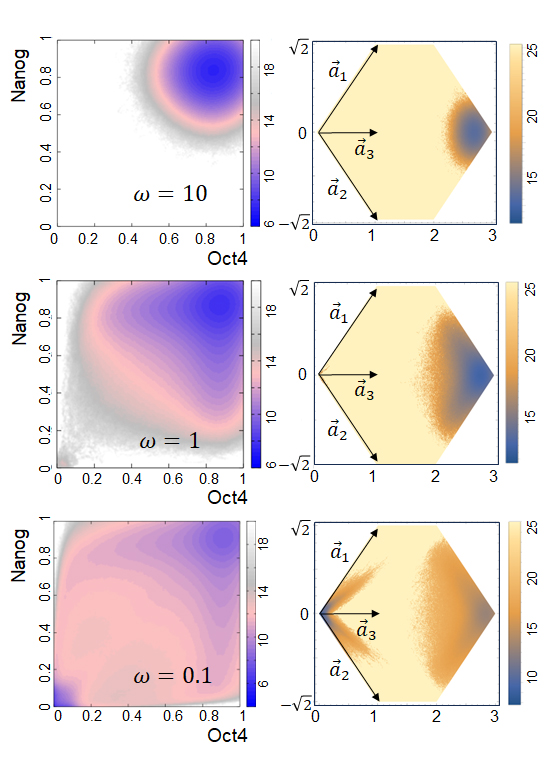}
\caption{
The landscape of the circuit of the three core genes of mES cells, \textit{Circuit C}, calculated with various values of the adiabaticity parameter $\omega=\omega_1=\omega_2=\omega_3$. The typical concentrations were set to $\bar{p}_1=\bar{p}_2=\bar{p}_3=2$. (Left) The two-dimensional  landscape $U(p_1, p_3)$ plotted on the plane of $p_1/\bar{p}_1$ (normalized Oct4 concentration) and $p_3/\bar{p}_3$ (normalized Nanog concentration). (Right) The three-dimensional landscape $U(p_1, p_2, p_3)$, with $p_2$ being the Sox2 concentration, projected on the two-dimensional plane using the coordinate $(p_1/\bar{p}_1)\vec{a}_1+(p_2/\bar{p}_2)\vec{a}_2+(p_3/\bar{p}_3)\vec{a}_3$ with $\vec{a}_1=(1,\sqrt{2})$, $\vec{a}_2=(1,-\sqrt{2})$, and $\vec{a}_3=(1,0)$.
}
\label{basins3}
\end{figure}

\subsection{Large fluctuations in the three-gene model}
We investigated a gene circuit composed of the three core genes in mES cells (\textit{Circuit C}). In this circuit, unlike the previous models \cite{Kalmar2009, Yu2018}, a direct self-activation of \textit{Nanog} is not considered, but the mutual activation of three genes, \textit{Oct4} \changed{($i=1$)}, \textit{Sox2} \changed{($i=2$)}, and \textit{Nanog} \changed{($i=3$)}, through the observed complex formation of TFs \cite{Chew2005, Masui2007, Niwa2007, Loh2006, Wang2006} is assumed. This gene circuit behaves similarly to \textit{Circuit A}, showing a similar dependence on the adiabaticity parameter $\omega$. We calculated a two-dimensional landscape $U(p_1,p_3)$ based on the projected distribution $P(p_1,p_3)$, where $p_1$ represents the concentration of Oct4 and $p_3$ represents the concentration of Nanog.  In Fig.\,\ref{basins3}, we show $U(p_1,p_3)$ for various values of adiabaticity, $\omega=\omega_1=\omega_2=\omega_3$. Additionally,  Fig.\,\ref{basins3} shows the projection of the cubic landscape $U(p_1,p_2,p_3)$  onto the two-dimensional plane to demonstrate the basin distribution in the three-dimensional space, with $p_2$ representing the concentration of Sox2.

When $\omega$ is large, the landscape has a single basin at the active state, and when $\omega$ is small, it has two basins at the active and inactive states. At a large value of $\omega= 10$, the active basin restricts fluctuations to a narrow region, while at a small value of $\omega= 0.1$, the landscape extends globally over the active and inactive basins, allowing large fluctuations in $p_1$, $p_2$, and $p_3$. Since $\omega$ controls the fluctuation amplitude, we propose a mechanism of using the heterogeneous $\omega_i$ to explain the experimentally observed heterogeneous fluctuations among the three core genes. 

\changed{
The adiabaticity $\omega_i$, is likely influenced by the timescale of chromatin structural reorganization. During cell differentiation, the chromatin structure surrounding the three core genes---\textit{Oct4}, \textit{Sox2}, and \textit{Nanog}---undergoes significant changes. In mES cells, the chromatin domains containing these core genes are part of a micrometer-sized structure known as the active compartment (A compartment). However, when differentiation begins, these domains transition to the inactive compartment (B compartment) \cite{deWit2013, Lando2024}.
It is plausible that even prior to differentiation,  fluctuations involving these domains in mES cells serve as precursors to compartment switching. Such large-scale conformational fluctuations are expected to occur over cell cycle periods. Moreover, these fluctuations involving the entire domain should correlate with domain-wide changes in collective histone modifications.
We propose that the transcription activity of \textit{Nanog} in mES cells is determined by this domain-wide switching, while the transcription activities of \textit{Oct4} and \textit{Sox2} are more significantly influenced by local acetylation and deacetylation around enhancers and promoters. Consequently, our hypothesis suggests that \textit{Nanog} exhibits slow transitions in its chromatin state, leading to an adiabaticity of $\omega_3 = 0.2 \sim 0.5$. In contrast, \textit{Oct4} and \textit{Sox2} demonstrate rapid transitions, resulting in greater adiabaticity of $\omega_1$ and $\omega_2$. 
While the reasons for this difference in adiabaticity among the different genes are not addressed here, we will proceed to discuss the model results to examine how this assumption shapes the landscapes of protein distributions.
 }

\begin{figure}[htbp]
\centering
\includegraphics[width=8cm]{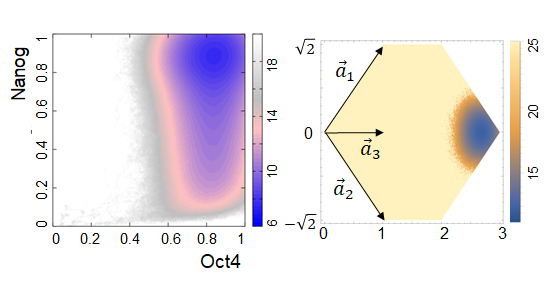}
\caption{
The landscape of the circuit of the three core genes of mES cells, \textit{Circuit C}, calculated with the heterogeneous adiabaticity parameters, $\omega_1=\omega_2=10$ for \textit{Oct4} and \textit{Sox2} and $\omega_3=0.5$ for \textit{Nanog}. Other parameters are the same as in Fig.\,\ref{basins3}. (Left) The two-dimensional  landscape $U(p_1, p_3)$. (Right) The three-dimensional landscape $U(p_1, p_2, p_3)$. The ways of plotting landscapes are the same as in Fig.\,\ref{basins3}.
}
\label{basins3hetero}
\end{figure}

\changed{
Fig.\,\ref{basins3hetero} shows the calculated landscape with heterogeneous adiabaticity parameters: $\omega_1=\omega_2=10$ for \textit{Oct4} and \textit{Sox2} and $\omega_3=0.5$  for \textit{Nanog}. The calculated landscape does not have a definite inactive basin. However, the distribution of the Nanog concentration $p_3$ is wide, indicating significant fluctuation in the \textit{Nanog} expression, while the expression level of \textit{Oct4} and \textit{Sox2}  shows narrow single-peak distributions. This is consistent with the experimentally observed heterogeneous fluctuations in mES cells. Thus, the timescale difference in the chromatin-state transitions allows the flexible tuning of fluctuations of individual genes in the circuit.
}

\changed{
In Appendix E, we present landscapes calculated using different parameterizations than the one used in Fig.\,\ref{basins3hetero}. 
Appendix E shows that it is crucial for the three genes to exhibit distinct differences in adiabaticity, highlighting the importance of regulatory mechanisms through variations in timescales.
}

\section{Deep epigenetic regulation}

 \changed{
In the previous section (Sec.\,IV), we demonstrated that the adiabatic switching of the chromatin states of  \textit{Oct4} and \textit{Sox2}, along with the nonadiabatic switching of the chromatin states of \textit{Nanog} creates the landscape that aligns with the experimentally observed heterogeneous fluctuations in mES cells. However, the underlying reason for the differences in adiabaticity among these three genes is not immediately clear. To address this issue and to test our model assumptions, it should be helpful to describe the chromatin states with additional degrees of freedom.
}

 \changed{
Gene transcription begins when the transcription initiation complex—composed of TFs, mediators, coactivators, RNA polymerase, and other molecules—assembles around the enhancer and promoter (EP) region of a gene. The EP region varies in size from 1\,kilobase (kb) to 100\,kb, depending on the characteristics of each gene. This region is a sub-structure within a chromatin domain, where the domain size typically ranges from 100\,kb to 1\,Mb in mammalian cells \cite{Misteli2020}. Observations of transcriptional bursting indicate that the transcription initiation complex forms and resolves near the EP over a timescale of $10^2$ to $10^3$\,seconds \cite{Fukaya2023, Porello2023, Meeussen2024}.
To capture these fluctuations, we introduce a variable $z_i$ to represent the EP state for the $i$th gene. We also use the variable $y_i$, inherited from the model in the previous section, to describe the collective histone methylation and demethylation throughout the entire domain. 
}

\begin{figure}[t]
\centering
\includegraphics[width=9cm]{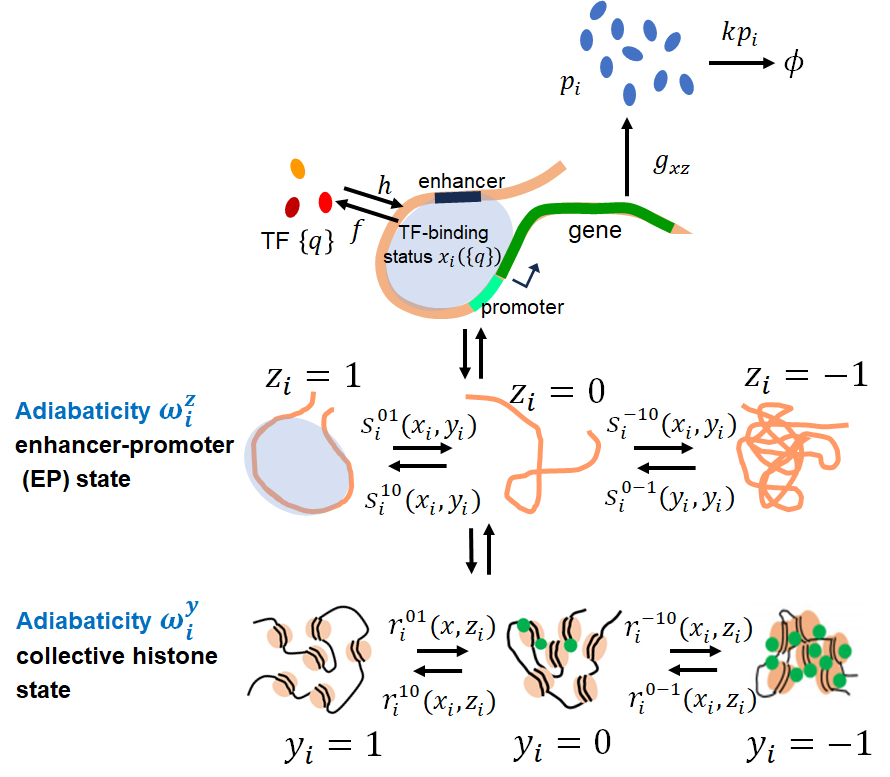}
\caption{Deep epigenetic regulation. 
\changed{
The model of deep epigenetic regulation describes multilayered regulatory mechanisms involving changes in three variables: $x_i$, $y_i$, and $z_i$. Here, $x_i$ represents the binding status of TFs, and $y_i$ indicates the collective histone methylation and demethylation across the domain. The variable $z_i$ represents the state of the enhancer-promoter (EP) region: $z_i = -1$ when the EP region near the $i$th gene is deacetylated and condensed. $z_i = 0$ when the EP region is acetylated and adopts an extended structure. $z_i = 1$ when the EP region is extended and forms a transcription initiation complex.
The rate of protein synthesis, denoted as $g_{xz}$, is affected by both $x_i$ and $z_i$. The value of $z_i$ can change at rates $s_i^{zz'}$, which depend on both $x_i$ and $y_i$, and the rates of collective histone-state transitions, $r_i^{yy'}$, are influenced by both $x_i$ and $z_i$.} 
}
\label{scheme2}
\end{figure}

A schematic representation of this model is shown in Fig.\,\ref{scheme2}. 
We define 
\changed{$z_i = -1$ when histones in the EP region are dacetylated making  the EP region condensed, $z_i = 0$ when they are acetylated and the EP region has an extended structure,}
 and $z_i = 1$ when the extended EP region is forming a transcription initiation complex. 
We consider that the protein-synthesis rate, $g(x_i,z_i)$, depends explicitly on the TF-binding status $x_i$ and the \changed{EP state} $z_i$, and indirectly on the collective histone state $y_i$ through the dependence of $z_i$ on $y_i$.
We define the transition rates $s_i^{zz'}$ and $r_i^{yy'}$ at the domain containing the $i$th gene;
transitions from the \changed{EP state} $z_i'$ to $z_i$ occur at the rate $s_i^{zz'}$, and transitions from the collective histone-state $y'_i$ to $y_i$ occur at the rate $r_i^{yy'}$.  We define parameters $\bar{s}^{zz'}$ and $\bar{r}^{yy'}$, which are of order of $k$, characterizing the transitions.
\changed{In the model discussed in the previous section (Sec. IV), we examined the scenario where the adiabaticity varies among different genes. In this section, we consider a model that accounts for variations in adiabaticity not only among genes but also between different layers of the chromatin's degrees of freedom in each gene. 
Thus, we introduce two types of }adiabaticity parameters, the \changed{EP adiabaticity} $\omega_i^z$ and the \changed{collective histone-state adiabaticity} $\omega_i^y$, and write 
\begin{eqnarray}
s_i^{zz'}&=&\omega_i^z \bar{s}^{zz'}, \nonumber \\
r_i^{yy'}&=&\omega_i^y \bar{r}^{yy'}.
\label{sr}
\end{eqnarray}
The parameters $\omega_i^z$ and $\omega_i^y $ measure the ratios of rates of $z$ and $y$ transitions over the rate of the $p$ fluctuation at the $i$th gene; Eq.\ref{sr} shows $s_i^{zz'}=O(\omega_i^z k)$ and $r_i^{yy'}=O(\omega_i^y k)$.

\changed{The EP state should be changed by the TF binding and the collective histone modification. 
Therefore, the EP-state transition rate $s_i^{zz'}$ and its normalized value are functions of $x_i$ and the collective histone state $y_i$ as $s_i^{zz'}(x_i,y_i)$ and $\bar{s}^{zz'}(x_i,y_i)$. 
The binding rate of the histone modifier enzymes should depend on the TF binding and the EP state  \cite{Bannister2011, Calo2013, Grewal2023}. Therefore,}
 the collective histone-state transition rate $r_i^{yy'}$ and its normalized value are functions of the TF state $x_i$ and the EP state $z_i$ as $r_i^{yy'}(x_i,z_i)$ and $\bar{r}^{yy'}(x_i,z_i)$. When we use the first-order approximation for them, we have
\begin{eqnarray}
\bar{s}^{zz'}(x_i,y_i)=\nu^{zz'}+\sigma_{x}^{zz'}x_i + \sigma_{y}^{zz'}y_{i+}, 
\label{Deepnu} \\ 
\bar{r}^{yy'}(x_i,z_i)=\mu^{yy'}+\gamma_{x}^{yy'}x_i + \gamma_{z}^{yy'}z_{i+}, 
\label{Deepmunu}
\end{eqnarray}
with $z_{i+}=(1+z_i)/2$. Here, $\nu^{zz'}$, $\sigma_{x}^{zz'}$, $\sigma_{y}^{zz'}$, $\mu^{yy'}$, $\gamma_{x}^{yy'}$, and $\gamma_{z}^{yy'}$ are the coefficients of the order of $k$. 
Thus, we have a multilayer regulation mechanism through \changed{sub-domain} transitions of $z$ and \changed{domain-wide collective histone-state} transitions of $y$ in chromatin. We 
refer to this multilayer regulation as \textit{deep epigenetic regulation} or \textit{deep epigenetics}.

Biochemical analyses, including studies on the complex cross-talks among histone modifier enzymes, have highlighted the relationships between \changed{the EP state, TF binding, and the collective histone modifications} \cite{Bannister2011, Calo2013, Grewal2023}. Although a comprehensive physical model of these relationships has not yet been established, a possible assumption is that changes in these states are consistent with one another. This consistency can be summarized as  $\sigma_{y}^{-10}\approx\sigma_{y}^{01}< 0$, $\sigma_{y}^{10}\approx \sigma_{y}^{0-1}> 0$, $\gamma_{z}^{-10}\approx\gamma_{z}^{01}< 0$, and $\gamma_{z}^{10}\approx\gamma_{z}^{0-1}> 0$. 
\changed{
If we were to apply different signs to these parameters, suggesting an inconsistency, this could result in negative feedback instead. Such time-delayed negative feedback might induce oscillatory behaviors, making it an intriguing area for the further exploration.
}

In the deep epigenetic regulation model, Eq.\,\ref{chroms} with $k=1$ is replaced by
\begin{eqnarray}
\frac{dp_i}{dt}&=&G_{p_i}-F_{p_i}+\eta_{p_i},  \nonumber \\
\frac{1}{\omega_i^z}\frac{dz_i}{dt}&=&G_{z_i}-F_{z_i} +\eta_{z_i}, \nonumber \\
\frac{1}{\omega_i^y}\frac{dv_i}{dt}&=&G_{y_i}-F_{y_i} +\eta_{y_i},
\label{Deepchroms}
\end{eqnarray}
with
$\left<\eta_{z_i}(t)\right>=\left<\eta_{p_i}(t)\eta_{z_j}(t')\right>=\left<\eta_{y_i}(t)\eta_{z_j}(t')\right>=0$, and
$\left<\eta_{z_i}(t)\eta_{z_j}(t')\right>=2D_{z_i}\delta_{ij}\delta(t-t')$, and Eq.\,\ref{diffusionC} becomes
\begin{eqnarray}
D_{p_i}&=&\frac{1}{2\Omega}\left(G_{p_i}+F_{p_i}\right), \nonumber \\
D_{z_i}&=&\frac{1}{2\omega_i^z}\left(G_{z_i}+F_{z_i}\right), \nonumber \\
D_{y_i}&=&\frac{1}{2\omega_i^y}\left(G_{y_i}+F_{y_i}\right).
\label{DeepdiffusionC}
\end{eqnarray}
In Eq.\,\ref{Deepchroms}, by writing $\xi(x,z)=g(x,z)/(k\Omega)$, $z_{i+}=(1+z_i)/2$, and $z_{i-}=(1-z_i)/2$, we have
\begin{eqnarray}
G_{p_i}&=&\left(\xi(1,1)z_{i+}^2+2\xi(1,0)z_{i+}z_{i-} + \xi(1,-1)z_{i-}^2\right)x_i  \nonumber \\
&+&\left(\xi(0,1)z_{i+}^2+2\xi(0,0)z_{i+}z_{i-} + \xi(0,-1)z_{i-}^2\right)(1-x_i),  \nonumber \\
G_{z_i}&=&\bar{s}^{0-1}(x_i,y_i)z_{i-}^2+2\bar{s}^{10}(x_i,y_i)z_{i+}z_{i-}, \nonumber \\
G_{y_i}&=&\bar{r}^{0-1}(x_i,z_i)y_{i-}^2+2\bar{r}^{10}(x_i,z_i)y_{i+}y_{i-}, \nonumber \\
F_{p_i}&=&p_i, \nonumber \\
F_{z_i}&=&2\bar{s}^{-10}(x_i,y_i)z_{ia}z_{i-}+\bar{s}^{01}(x_i,y_i)z_{i+}^2, \nonumber \\
F_{y_i}&=&2\bar{r}^{-10}(x_i,z_i)y_{i+}y_{i-}+\bar{r}^{01}(x_i,z_i)y_{i+}^2.
\label{DeepforceC}
\end{eqnarray}

\begin{figure}[htbp]
\centering
\includegraphics[width=8cm]{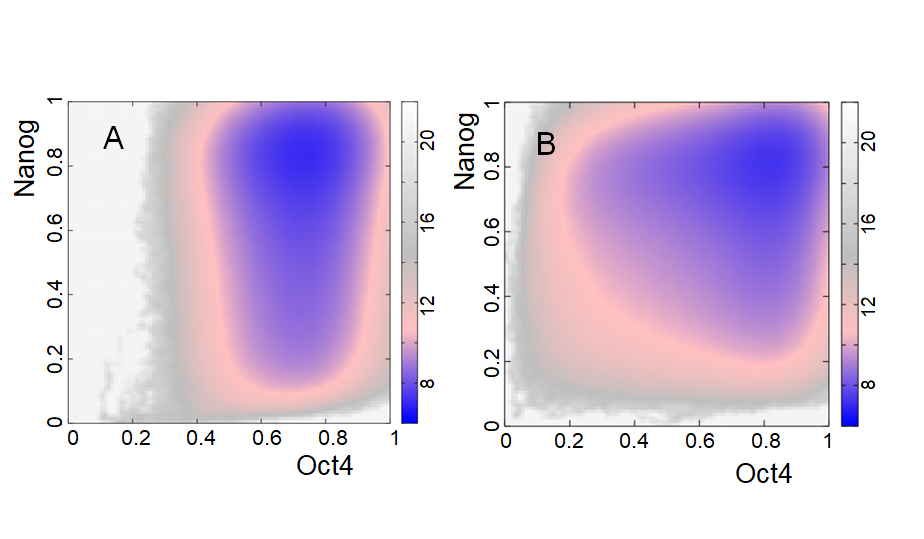}
\caption{
\changed{The two-dimensional landscape, $U(p_1, p_3)$, of the circuit of the three core genes of mES cells, \textit{Circuit C}, calculated with the deep epigenetic regulation model. $\bar{p}=2$ and other parameters are from Table III. (A) The collective histone-state adiabaticity is $\omega_1^y=\omega_2^y=\omega_3^y=0.1$. The EP adiabaticity is  $\omega_1^z=\omega_2^z=10$ for \textit{Oct4} and \textit{Sox2}, and $\omega_3^z=0.5$ for \textit{Nanog}. (B) The collective histone-state adiabaticity is $\omega_1^y = \omega_2^y = 10$ for \textit{Oct4} and \textit{Sox2}, and $\omega_3^y = 0.5$ for \textit{Nanog}. The EP adiabaticity is $\omega_1^z = \omega_2^z = \omega_3^z = 1$.
}
}
\label{DeepLand}
\end{figure}

\changed{
We use the parameter values from Table III to numerically integrate Eqs.\,\ref{Deepchroms}--\ref{DeepforceC}. Since the timescale of collective histone-state modifications spans across cell cycles, we expect the collective histone-state adiabaticity, $\omega_i^y$, to be low and nonadiabatic. In contrast, the EP state represents a local region of the domain, suggesting that the EP adiabaticity, $\omega_i^z$, should generally be high.
However, we propose a hypothesis that the transcription initiation complex of \textit{Nanog} spans domain-wise, thereby linking the EP state of \textit{Nanog} to the collective histone state of the entire domain. This connection results in a nonadiabatic behavior for \textit{Nanog}, characterized by a low value of $\omega_3^z$.}

\begin{figure}[htbp]
\centering
\includegraphics[width=9cm]{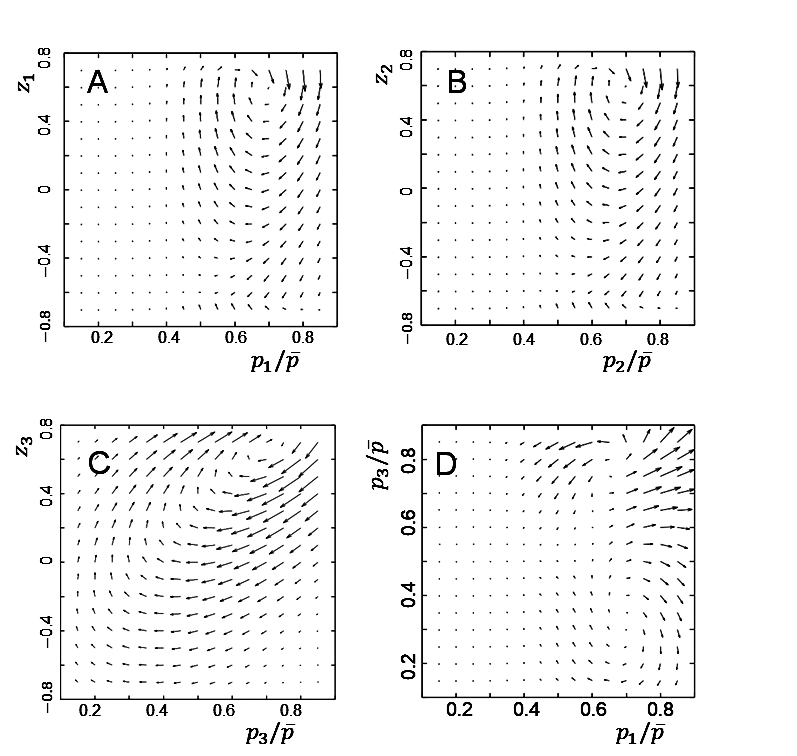}
\caption{
\changed{
The probability current $\vec{J}$ in the deep epigenetic regulation model of the circuit of three core genes of mES cells, \textit{Circuit C}, projected on the two-dimensional planes. (A) $\vec{J}=(J_{p_1}, J_{z_1})$ projected on the plane of the normalized concentration of Oct4, $p_1/\bar{p}$, and the EP state of \textit{Oct4}, $z_1$. (B) $\vec{J}=(J_{p_2}, J_{z_2})$ projected on the plane of the normalized concentration of Sox2, $p_2/\bar{p}$, and the EP state of \textit{Sox2}, $z_2$. (C) $\vec{J}=(J_{p_3}, J_{z_3})$ projected on the plane of the normalized concentration of Nanog, $p_3/\bar{p}$, and the EP state of \textit{Nanog}, $z_3$. (D) $\vec{J}=(J_{p_1}, J_{p_3})$ projected on the plane of the normalized concentration of Oct4, $p_1/\bar{p}$, and the normalized concentration of Nanog, $p_3/\bar{p}$.
 The parameters are the same as in Fig.\,\ref{DeepLand}A.
}
}
\label{NanogCurr}
\end{figure}

\changed{
In Fig.\,\ref{DeepLand}A, we present the landscape calculated using the adiabaticity parameters outlined above, with small values for $\omega_i^y$ and heterogeneous values for $\omega_i^z$: specifically, we used $\omega_1^y=\omega_2^y=\omega_3^y=0.1$, $\omega_1^z=\omega_2^z=10$, and $\omega_3^z=0.5$. This landscape is similar to the one shown in Fig.\,\ref{basins3hetero} and aligns with the experimentally observed heterogeneous fluctuations. It is important to note that the protein-production rate $g(x_i, z_i)$ directly depends on the TF-binding state $x_i$ and the EP state $z_i$  in our current model. In contrast, the collective histone state $y_i$ only indirectly influences $g(x_i, z_i)$ through the dependence of $z_i$ on $y_i$. As a result,  the combination of heterogeneous collective histone state adiabaticity $\omega_i^y$ and homogeneous EP adiabaticity $\omega_i^z$ produces a landscape characterized by homogeneous fluctuations across the three genes, which is inconsistent with the experimentally observed heterogeneous fluctuations of these genes. Fig.\,\ref{DeepLand}B shows an example of such a landscape calculated using the adiabaticity parameters, $\omega_1^y = \omega_2^y = 10$, $\omega_3^y = 0.5$, and $\omega_1^z = \omega_2^z = \omega_3^z = 1$.
}

\changed{
The probability currents calculated with this model suggest a way to check our hypothesis on the timescale separation in chromatin regulation. Figs.\,\ref{NanogCurr}A--\ref{NanogCurr}C show the probability currents calculated with the deep epigenetic regulation model using the parameters of Fig.\,\ref{DeepLand}A. These currents are projected onto the $p_i$-$z_i$ plane for \textit{Oct4} ($i=1$, Fig.\,\ref{NanogCurr}A), \textit{Sox2}  ($i=2$, Fig.\,\ref{NanogCurr}B), and \textit{Nanog} ($i=3$, Fig.\,\ref{NanogCurr}C). Figs.\,\ref{NanogCurr}A and \ref{NanogCurr}B illustrate the currents moving parallel to the $z_i$ direction, highlighting the rapid changes in the EP state of \textit{Oct4} and \textit{Sox2}. In contrast, \textit{Nanog} exhibits a circular flow that rotates in a diagonal direction (Fig.\,\ref{NanogCurr}C), indicating correlated fluctuations between the EP state and protein synthesis, along with hysteresis in the fluctuations of \textit{Nanog} within mES cells. In the full-dimensional space, the probability current flows in a circular manner and has no divergence in a steady state. However, when we project this current onto the two-dimensional plane representing the concentrations of Oct4 ($p_1$) and Nanog ($p_3$), we observe divergence occurring from the high-Nanog state (Fig.\,\ref{NanogCurr}D). Moreover, this projected current moves in opposite directions when comparing the high-Nanog state to the low-Nanog state  (Fig.\,\ref{NanogCurr}D). We anticipate that experimentally verifying these predicted probability flows by measuring dynamic fluctuations in mES cells will enhance our understanding and help us test our hypothesis.
}

\section{Discussion}
We studied models of gene circuits made up of one to three interacting eukaryotic genes. We described the speed of changes in the chromatin state using the adiabaticity parameter  $\omega$, and investigated the stochastic behaviors of the model by varying $\omega$ using the landscape picture. The landscape has different basin structures in the adiabatic ($\omega\gg 1$) and nonadiabatic ($\omega \ll 1$) limits, and it shows enhanced circular flow of the probability current in between the two limits, known as the eddy regime ($0.1\lesssim \omega\lesssim 1$). \changed{Entropy production---which serves as a measure to assess how system deviates from the equilibrium---has a large value in this eddy regime.}

\changed{
Eukaryotic genes are regulated by a variety of interconnected processes within chromatin. Among these processes, the deep epigenetic regulation model discussed in Sec.\,V focuses on two key aspects: changes in the enhancer-promoter region, which is a local structure within the chromatin domain, and the collective histone methylation and demethylation across the entire domain. This framework opens a way to compare the calculated results and experimental data related to these chromatin degrees of freedom. This model can be expanded to incorporate further coexisting regulatory mechanisms in chromatin, such as cross-talks among multiple histone modifications, DNA methylation, and the formation and resolution of chromatin domains. With such extension to include multiple degrees of freedom, we can draw parallels between  deep epigenetic regulation and information processing in artificial neural networks. Each gene receives a set of TFs as inputs, produces a protein as output, and the interconnected and multilayered regulations within its chromatin domain can be viewed as hidden layers of regulation. We anticipate that  these hidden layers enhance regulatory flexibility, similar to how neural networks improve their performance, as discussed in the context of the universal approximation theorem regarding representation capability  \cite{Tikk2003}. Additionally, the couplings of variables $x_i$, $y_i$, and $z_i$ mentioned in Eq.\,\ref{DeepforceC} are multiplicative, much like the multiplicative gating used in recurrent neural networks, which has been shown to improve network efficiency \cite{Krishnamurthy2022}. Exploring these parallels between chromatin regulation and efficient information processing in neural networks presents an intriguing direction for future studies, as these insights could enhance our understanding of the evolutionary origins of various overlapping regulatory mechanisms within chromatin domains.
}

\changed{Timescales within cells can be influenced by changes in the enzymatic activities involved in gene regulation processes. Such changes may lead to transitions between different cell types. This concept can be illustrated by model circuits. In the nonadiabatic regime of $\omega\ll 0.1$ for \textit{Circuit B}, for instance, a basin characterized by a high value of $p_1$ or $p_2$ is distinct, as shown in Fig.\,\ref{basins2}, stabilizing the cells residing there. However, if the chromatin-state change is accelerated, shortening the timescale, the system enters to the eddy regime of $0.1\lesssim \omega\lesssim1$. In this regime, the basins become shallower, causing the cells to fluctuate across a wide range of $p_1$ and $p_2$ (Fig.\,\ref{basins2}). If the timescale is then increased again, bringing the system back to the nonadiabatic regime, the basins with high $p_1$ or $p_2$ become stabilized once more. As a result, the fluctuating cells may converge into a different basin than the one they originally occupied, leading to a transition in cell type. Large fluctuations are expected during this transition, as seen in various models of gene regulation \cite{Chalancon2012}. The current model further predicts that there will be increased circular flow during the transition in the eddy regime. This flow should be experimentally observed as a correlation between fluctuations in protein concentrations and variations in chromatin states during the cell-type transition.}

The circular flow over the extended landscape indicates the presence of hysteresis during transitions. This observation suggests a temporal sequence in which chromatin-state changes precede changes in expression levels. This time ordering was indeed noted during the differentiation process of mouse embryonic cells \cite{Liu2023}. While the traditional biochemical perspective may focus on specific proteins or genes that regulate this timing, our model proposes that the time ordering results not from specific regulatory factors in cells, but rather from the interaction between slow and fast processes in gene regulation. In this framework, the slow process governs the system's state transitions, while the fast process reacts to it. This concept, where the slow process determines the system's fate, is similar to findings from simulations of a multilayer network model \cite{Nicoletti2024}. The mechanism proposed in this study can be validated if the adiabaticity parameter $\omega$ can be controlled in cells.

\begin{figure}[htbp]
\centering
\includegraphics[width=9cm]{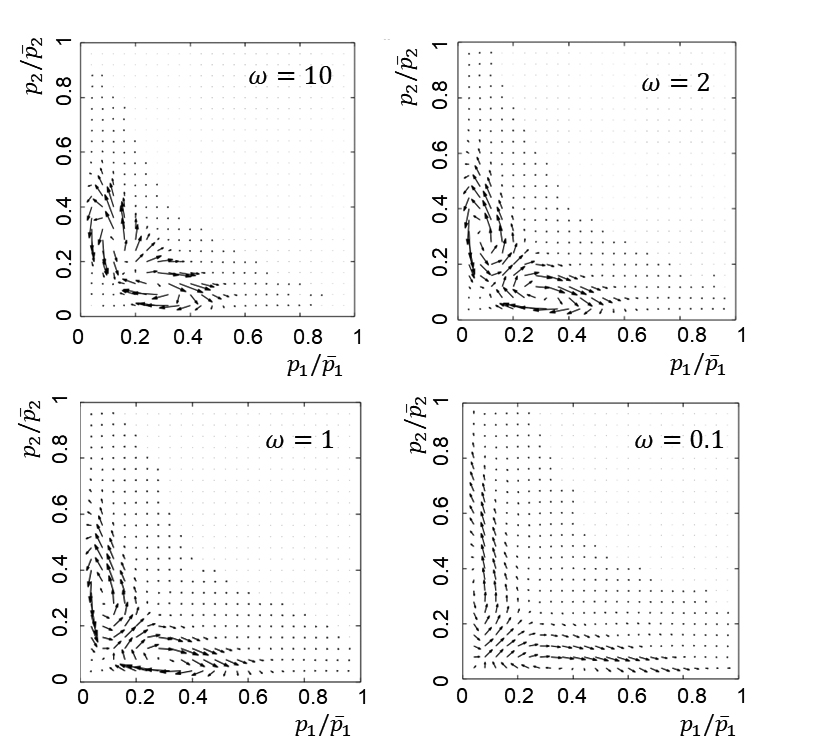}
\caption{
The probability current $\vec{J}=(J_{p_1}, J_{p_2})$ of the mutually repressing two-gene circuit, \textit{Circuit B}, projected on the two-dimensional $p_1$-$p_2$ plane. The parameters are the same as in Fig.\,\ref{basins2}.
}
\label{precursor}
\end{figure}

Another feature of the circular flux is found in the example of the mutually repressing two-gene circuit (\textit{Circuit B}). This circuit shows a bifurcation of a basin at the diagonal position to a pair of basins at off-diagonal positions in the eddy regime. This bifurcation takes place at $\omega=\omega_\textrm{c}\approx 1.2$ with parameters used in Fig.\,\ref{basins2}. Fig.\,\ref{precursor} shows the probability current plotted on the $p_1$-$p_2$ plane, showing the existence of a pair of circular fluxes rotating in opposite directions around positions of a pair of off-diagonal basins. We find that a pair of fluxes appear even in the case of $\omega>\omega_\textrm{c}$, showing that circular fluxes are a precursor, or a warning sign \cite{Masuda2024}, of the bifurcation. It is intriguing to check this prediction by the RNA-velocity measurement in cells undergoing the switching fluctuations.

A crucial area for future research is to conduct more direct comparisons of the simulation with experimental data. By employing combined high-throughput methodologies such as Hi-C, ChIP-seq, and RNA-seq, it is becoming possible to infer the landscapes and probability currents of gene regulation in cells \cite{Maehara2019, Qiu2022, Su2024, Zhu2024}. Building landscapes and analyzing probability currents from experiments will enable comparisons with the theoretical methods, paving the way for further exploration into the biological physics of gene regulation.

\section*{Acknowledgements}
This work was supported by JSPS-KAKENHI Grants 22H00406 and 24H00061.

\section*{Appendices}

\begin{table*}
 \begin{center}
   \caption{Parameters used in each circuit}
\begin{tabular}{ccrrr}  \hline
Parameters & & \textit{Circuit A} & \textit{Circuit B} & \textit{Circuit C} \\ \hline 
 \hline
\footnotesize{Coefficients used in }
&$\mu^{0-1}$ & $k$ & $k$ & $k$ \\

$\bar{r}_i^{yy'}(x_i)=\mu^{yy'}+\gamma^{yy'}x_i$\footnotesize{(Eq.\,\ref{munu})} 
&$\mu^{10}$ & $k$ & $k$ & $2k$ \\ 

\footnotesize{determining the transition rate} 
&$\mu^{01}$ & $k$ & $k$ & $k$ \\ 

\footnotesize{from a chromatin state $y'$ to the other $y$}~~~~~~~~
&$\mu^{-10}$ & $k$ & $k$ & $2k$ \\ 

 &$\gamma^{0-1}$ & $0.6k$ & $-0.6k$ & $k$ \\ 

 &$\gamma^{10}$ & $0.6k$ & $-0.6k$ & $k$ \\ 

 &$\gamma^{01}$ & $-0.6k$ & $0.6k$ & $-k$ \\ 

 &$\gamma^{-10}$ & $-0.6k$ & $0.6k$ & $-k$ \\ 
\hline

\footnotesize{TF binding/unbinding parameter ratio} & $h_0/f$  & 10  & 10 & 10 \\
  & $h_1/f$  & - & - & 200 \\
\hline

\footnotesize{Normalized protein-synthesis rate}  &$\xi_{11}$ & $\bar{p}$ & 0.2 & $\bar{p}$ \\
$\xi_{xy}=g_{xy}/(k\Omega)$ &$\xi_{01}$ & 0.2 & $\bar{p}$ & 0.2 \\
\footnotesize{at the TF-binding state $x$} &$\xi_{10}$ & 0.2 & 0 & 0.5\\
\footnotesize{and the chromatin state $y$} &$\xi_{00}$ & 0 & 0.2 & 0 \\
&$\xi_{1-1}$ & 0 & 0 & 0 \\
&$\xi_{0-1}$ & 0 & 0 & 0 \\ 
\hline

\label{table:each_circuit}
\end{tabular}
 \end{center}
\end{table*}

\begin{table*}
 \begin{center}
   \caption{Parameters in the deep epigenetic regulation model}
\begin{tabular}{ccrccr}  \hline

\footnotesize{Coefficients used in } & 
$\nu^{0-1}$ & $k$ &
\footnotesize{Coefficients used in } &
$\mu^{0-1}$ & $k$ \\

~~~$\bar{s}^{zz'}=\nu^{zz'}+\sigma_{x}^{zz'}x_i+ \sigma_{y}^{zz'}y_{i+}$~\footnotesize{(Eq.\,\ref{Deepnu})}~~& 
$\nu^{10}$ & $2k$ &
~~~$\bar{r}^{yy'}=\mu^{yy'}+\gamma_{x}^{yy'}x_i+ \gamma_{z}^{yy'}z_{i+}$~\footnotesize{(Eq.\,\ref{Deepmunu})}~~&
$\mu^{10}$ & $2k$ \\

 \footnotesize{determining the transition rate from }& 
$\nu^{01}$ & $k$ &
 \footnotesize{determining the transition rate from }&
$\mu^{01}$ & $k$ \\

 \footnotesize{an enhancer-promoter state $z'$ to the other $z$}~~& 
$\nu^{-10}$ & $2k$ &
 \footnotesize{~~~a collective histone state $y'$ to the other $y$~~}&
$\mu^{-10}$ & $2k$ \\ \\

& 
 $\sigma_x^{0-1}$ & $0.5k$ & 
&
 $\gamma_x^{0-1}$ & $0.2k$  \\

& 
 $\sigma_x^{10}$ & $0.5k$ & 
&
 $\gamma_x^{10}$ & $0.2k$  \\

& 
 $\sigma_x^{01}$ & $-0.5k$ & 
&
 $\gamma_x^{01}$ & $-0.2k$  \\

 & 
 $\sigma_x^{-10}$ & $-0.5k$ & 
 &
 $\gamma_x^{-10}$ & $-0.2k$  \\
\\

 & 
 $\sigma_y^{0-1}$ & $0.5k$ & 
 &
 $\gamma_z^{0-1}$ & $0.8k$  \\

 & 
 $\sigma_y^{10}$ & $0.5k$ & 
 &
 $\gamma_z^{10}$ & $0.8k$  \\

 & 
 $\sigma_y^{01}$ & $-0.5k$ & 
 &
 $\gamma_z^{01}$ & $-0.8k$  \\

 & 
 $\sigma_y^{-10}$ & $-0.5k$ & 
 &
 $\gamma_z^{-10}$ & $-0.8k$  \\
  \hline

\footnotesize{TF binding/unbinding parameter ratio}
& $h_0/f$ & 10 &  &  &\\
& $h_1/f$ & 200 &  &  &\\
\hline

\end{tabular}

\begin{tabular}{ccrlrlr}
~~~~~~~~~~~~\footnotesize{Normalized protein-synthesis rate}~~~~~~~~~~~~~~~~~~~~~~~~~
&$\xi(1,1)$ & $\bar{p}$~~~~~~~~~~~~ 
& $\xi(1,0)$ & 0.2~~~~~~~~~~~~  
& $\xi(1,-1)$ & 0 \\

~~~~~~~~~~~~$\xi(x,z)=g(x,z)/(k\Omega)$ ~~~~~~~~~~~~~~~~~~~~~~~~~~~~~~~~
&$\xi(0,1)$ & 0.5~~~~~~~~~~~~ 
& $\xi(0,0)$  & 0~~~~~~~~~~~~
& $\xi(0,-1)$ & 0 \\
\hline
\end{tabular}

\label{table:epigenetic}
 \end{center}
\end{table*}

\subsection*{A: Numerical simulations}
Eq.\,\ref{chroms} and Eq.\,\ref{Deepchroms} were numerically solved  by the Euler discretization with a time step of $\delta t=0.01$ for $\omega<0.5$, $\delta t=0.001$ for $0.5\le \omega <10$, and $\delta t=0.0001$ for $10\le \omega$. In the discretized calculation, occasional large noise values in Eq.\,\ref{chroms} and Eq.\,\ref{Deepchroms} caused rare jumps, taking the trajectory out of the physically plausible ranges of  $0\le p_i$, $-1\le y_i \le 1$, and $-1\le z_i\le 1$. To address this issue, solid walls were assumed at the boundaries of these ranges in the numerical calculations.  The area near these boundary walls (5\% of the total area) was excluded in the integration in Eq.\,\ref{EPR} to prevent the artifact contribution of the reflecting probability flow from the walls.

\subsection*{B: Circuits}
\subsubsection*{Circuit A, Self-activating single-gene circuit}
The simplest circuit is a loop involving a single gene that produces a TF which activates itself   (Fig.\,\ref{circuits}A). The self-activation is ubiquitous in cells \cite{Alon2007}, and this circuit is the same as analyzed in the previous report \cite{Bhaswati2020}. In the present study, we investigate this circuit in greater depth and compare the dynamics with entropy production and time asymmetry in cross-correlation, which measures how far the system is from equilibrium. 
 
 We assume that the protein dimerizes to become a TF, and this dimerization occurs rapidly enough compared to the rate $k$ that we can express the TF concentration $q$ as $q \propto p^2$. Then, we write $h=h_0p^2$ with a constant $h_0>0$. We use the adiabatic approximation for representing the TF-binding state $x$ with its equilibrium value $x=h_0p^2/(h_0p^2+f)$.  We write $\bar{p}=\xi_{11}$, which is a typical value of $p$. 

\subsubsection*{Circuit B, Mutually repressing two-gene circuit}
We consider a circuit of mutually repressing two genes (Fig.\,\ref{circuits}B). Though this mutually repressing circuit is ubiquitous in eukaryotic cells, including the synthetically designed circuit \cite{Zhou2023}, we here consider a simplified model to focus on eddy dynamics of this type of circuit by explicitly analyzing the dynamics of the chromatin-state variables $y_1$ and $y_2$ in genes 1 and 2, respectively. 

To highlight the effects of eddy dynamics of the circuit, we assume that two genes have the same parameters; the adiabaticity is measured by the adiabaticity parameter $\omega=\omega_1=\omega_2$. We consider the case where dimers of protein produced from genes 1 and 2 become TFs binding on genes 2 and 1, respectively, so that  $q_1\propto p_1^2$ and $q_2\propto p_2^2$, leading to $h^{12}=h_0p_2^2$ and $h^{21}=h_0p_1^2$. We use the adiabatic approximation to derive $x_1$ and $x_2$, resulting in $x_1=x^{12}=h_0p_2^2/(h_0p_2^2+f)$ and $x_2=x^{21}=h_0p_1^2/(h_0p_1^2+f)$.
We write $\bar{p}=\xi_{10}$. 

\subsubsection*{Circuit C, Three-gene circuit model of pluripotency}
By exploring \textit{Circuits A} and \textit{B}, we analyze how the landscapes and nonequilibrium fluctuations depend on the adiabaticity $\omega_i$ of the chromatin-state dynamics. As an example biological system to apply these concepts of landscapes and nonequilibrium fluctuations, we consider the gene circuit to maintain pluripotency in mES cells composed of mutually activating three genes,  \textit{Oct4} ($i=1$), \textit{Sox2} ($i=2$), and \textit{Nanog} ($i=3$)  \cite{Boyer2006, Niwa2007} (Fig.\,\ref{circuits}C). A complex of Oct4 and Sox2 activates all three genes \cite{Chew2005, Masui2007, Niwa2007} and a complex of Oct4 and the Nanog-dimer activates \textit{Oct4} and \textit{Nanog} \cite{Loh2006, Wang2006}. We represent these regulations as 
\begin{eqnarray}
x_1&=&x_3=\frac{h_0p_1p_2}{h_0p_1p_2+f_0}\frac{h_1p_1p_3^2}{h_1p_1p_3^2+f_1}, \nonumber \\
x_2&=&\frac{h_0p_1p_2}{h_0p_1p_2+f_0}.
\end{eqnarray}
We use $\omega=\omega_1=\omega_2=\omega_3$ at the beginning, but we assume the heterogeneous $\omega_i$ later to compare the simulated data with the experimental results. For simplicity, we assume that the other parameters do not depend on $i$. We write $\bar{p}=\xi_{11}$.

\subsection*{C: Monomer TFs}

\begin{figure}[htbp]
\centering
\includegraphics[width=9cm]{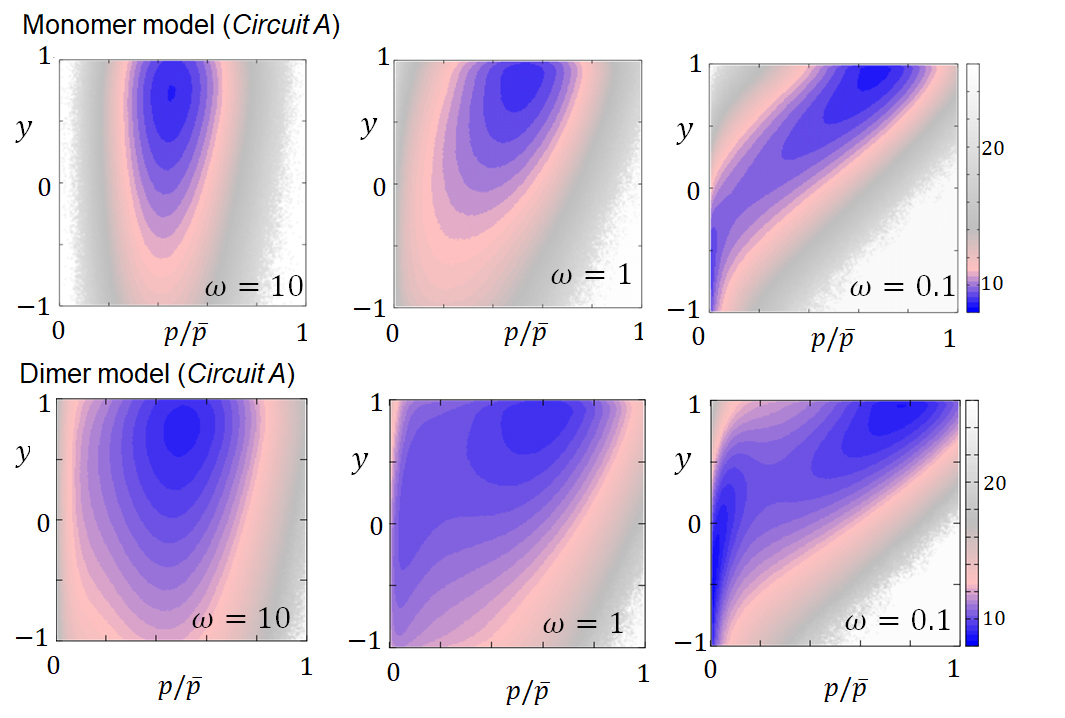}
\caption{
\changed{The landscape, $U(p,y)$, of the self-activating single-gene circuit (\textit{Circuit A}) is presented for two different models: one with a monomer TF where $ h_\text{m}=0.4h_0$ (Top) and another with a dimer TF where $h_0/f = 10$ (Bottom). The landscapes are compared by varying the adiabaticity $ \omega$. All other parameters remain same among panels with $\bar{p}=1$. The bottom panels are identical to those in Fig. \ref{basins1}.
}
}
\label{monomerA}
\end{figure}

\begin{figure}[htbp]
\centering
\includegraphics[width=8cm]{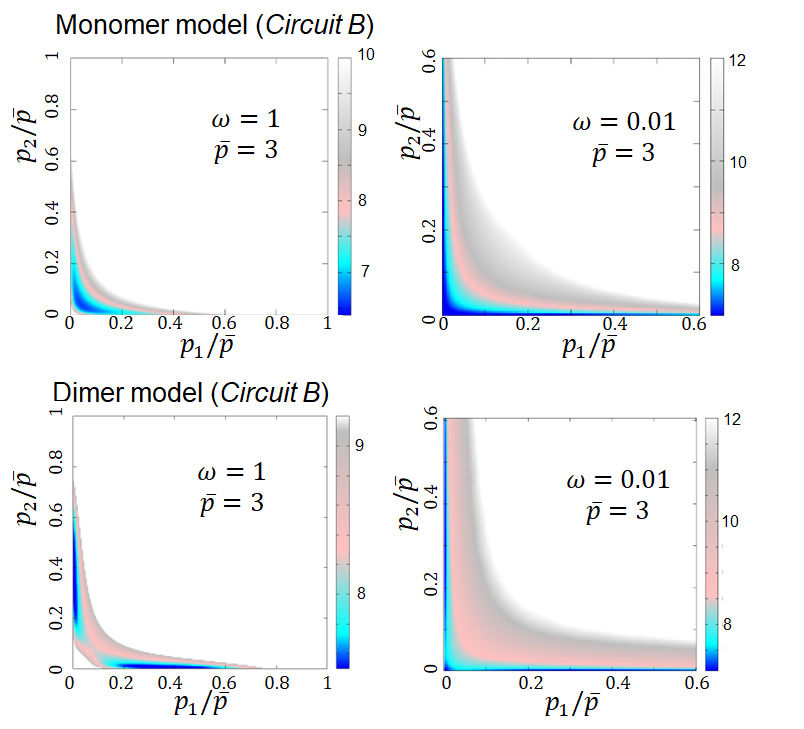}
\caption{
\changed{The landscape, $U(p_1,p_2)$, of the mutually repressing two-gene circuit (\textit{Circuit B}) is presented for two different models: one with monomer TFs where $ h_\text{m}=1.2h_0$ (Top) and another with dimer TFs where $h_0/f = 10$ (Bottom). The landscapes are compared by varying the adiabaticity $ \omega$. All other parameters remain same among panels with $\bar{p}=3$. 
}
}
\label{monomerBp3}
\end{figure}

\begin{figure}[htbp]
\centering
\includegraphics[width=8cm]{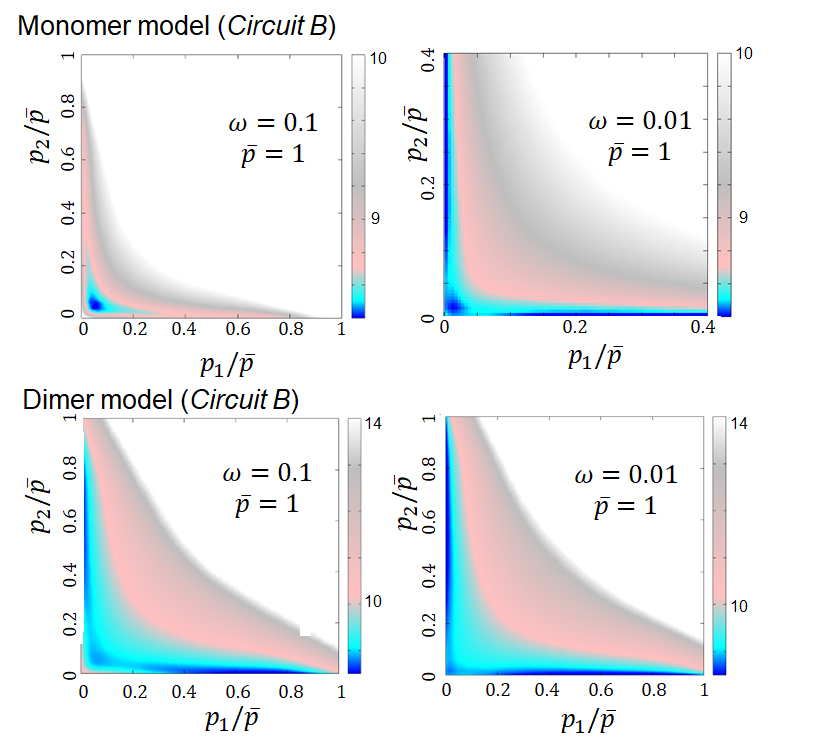}
\caption{
\changed{The landscape, $U(p_1,p_2)$, of the mutually repressing two-gene circuit (\textit{Circuit B}) is presented for two different models: one with monomer TFs where $ h_\text{m}=0.4h_0$ (Top) and another with dimer TFs where $h_0/f = 10$ (Bottom). The landscapes are compared by varying the adiabaticity $ \omega$. All other parameters remain same among panels with $\bar{p}=1$. 
}
}
\label{monomerBp1}
\end{figure}

\changed{In the model described in the main text, we assumed that TFs exist as dimers of the corresponding product proteins for describing \textit{Circuits A} and \textit{B} as explained in Appendix B. Specifically, we set  $h=h_0p^2$ in \textit{Circuit A}, and  $h^{12}=h_0p_2^2$ and $h^{21}=h_0p_1^2$ in \textit{Circuit B}. This dimer assumption emphasizes the nonlinear effects of TF binding and unbinding. In contrast, if we adopt a monomer assumption with $h=h_\text{m}p$ in \textit{Circuit A} and  $h^{12}=h_\text{m}p_2$ and $h^{21}=h_\text{m}p_1$ in \textit{Circuit B}, the nonlinear effects are diminished, leading to a weaker tendency for the coexistence of multiple basins. This reduced tendency can be observed in the landscapes derived from the monomer model (Figs.\,\ref{monomerA}--\ref{monomerBp1}). 
To facilitate a straightforward comparison between the dimer and monomer models, we set the parameter $h_\text{m}$ to $h_\text{m} = h_0 p_0$. Here, the typical protein concentration $p_0$ is defined as $p_0 = \alpha \bar{p}$ with $0 < \alpha < 1$. In this Appendix (Appendix C), we choose $\alpha = 0.4$, which gives us $h_\text{m} = 0.4 h_0 \bar{p}$. Specifically, when $\bar{p} = 3$, we have $h_\text{m} = 1.2 h_0$, and when $ \bar{p} = 1$, $h_\text{m} = 0.4 h_0$.
}

\changed{Fig.\,\ref{monomerA} compares the monomer and dimer models for the self-activating single-gene circuit (\textit{Circuit A}) by varying the chromatin adiabaticity $\omega$.  In both models, when the circuit is in the adiabatic regime at $\omega=10$, a single basin is observed at the active state of the chromatin, characterized by a high $y$ value. In the dimer model, when $\omega\approx 0.5$ in the eddy regime, an additional basin appears at the inactive state with a low $y$ value. A precursor of this new basin can be seen as a broadened basin in the case of $\omega=1$. However, in the monomer model, there is no indication of the emergence of the inactive basin at $\omega=1$. At $\omega=0.1$, both the dimer and monomer models exhibit two coexisting basins representing inactive and active states, but the inactive basin in the monomer model remains shallow.
}

\changed{Fig.\,\ref{monomerBp3} compares the monomer and dimer models for the mutually repressing two-gene circuit (\textit{Circuit B}) at a relatively high protein production level of $\bar{p} = 3$. At this high level of protein production, the dimer model displays two coexisting basins even at $\omega = 1$. In contrast, the monomer model exhibits a single basin in the inactive state, located at the diagonal position in the landscape.
In the nonadiabatic regime, with $\omega = 0.01$, the dimer model exhibits three basins. Meanwhile, the basin in the monomer model extends along the axes of $p_1 \approx 0$ and $p_2 \approx 0$, indicating a tendency to reach the two off-diagonal states. However, the monomer model's landscape does not yield multiple disconnected basins.
}

\changed{
Fig.\,\ref{monomerBp1} shows landscapes of the mutually repressing two-gene circuit (\textit{Circuit B}) at a lower protein production level $\bar{p} = 1$. The dimer model shows three coexisting basins at $\omega=0.1$. In contrast, the monomer model displays only a single basin. However, in the nonadiabatic regime at $\omega=0.01$, both the monomer and dimer models exhibit similar landscapes, with three basins in each. 
}

\changed{In summary, the diminished nonlinearity in the monomer model suppresses the tendency for forming distinct multiple basins.
However, in both the monomer and dimer models, decreasing the adiabaticity of chromatin-state transitions broadens the distribution of states, thereby promoting the emergence of multiple states. 
}

\subsection*{D: Two-state chromatin transitions}

\changed{In the main text, we described the chromatin-state transitions as  transitions between active ($y_i=1$) and inactive ($y_i=-1$) states through an intermediate state ($y_i=0$), as illustrated in Fig.\,\ref{scheme}. The significance of this intermediate state was highlighted by the nonzero parameters $\xi_{10}$, $\xi_{00}$, $\bar{r}^{10}$, and $\bar{r}^{-10}$ in Eq.\,\ref{forceC}. We can reduce this emphasis by setting $\xi_{10}=\xi_{00}=\bar{r}^{10}=\bar{r}^{-10}=0$. In this case, Eq.\,\ref{forceC} simplifies to
\begin{eqnarray}
G_{p_i}&=&\left(\xi_{11}^*y_{i+}^2 + \xi_{1-1}^*y_{i-}^2\right)x_i  +\left(\xi_{01}^*y_{i+}^2+ \xi_{0-1}^*y_{i-}^2\right)(1-x_i),  \nonumber \\
G_{y_i}&=&\bar{r}^{1-1}y_{i-}^2/k, \nonumber \\
F_{p_i}&=&p_i, ~~~
F_{y_i}=\bar{r}^{-11}y_{i+}^2/k,
\label{forceC2}
\end{eqnarray}
}

\begin{figure}[htbp]
\centering
\includegraphics[width=8cm]{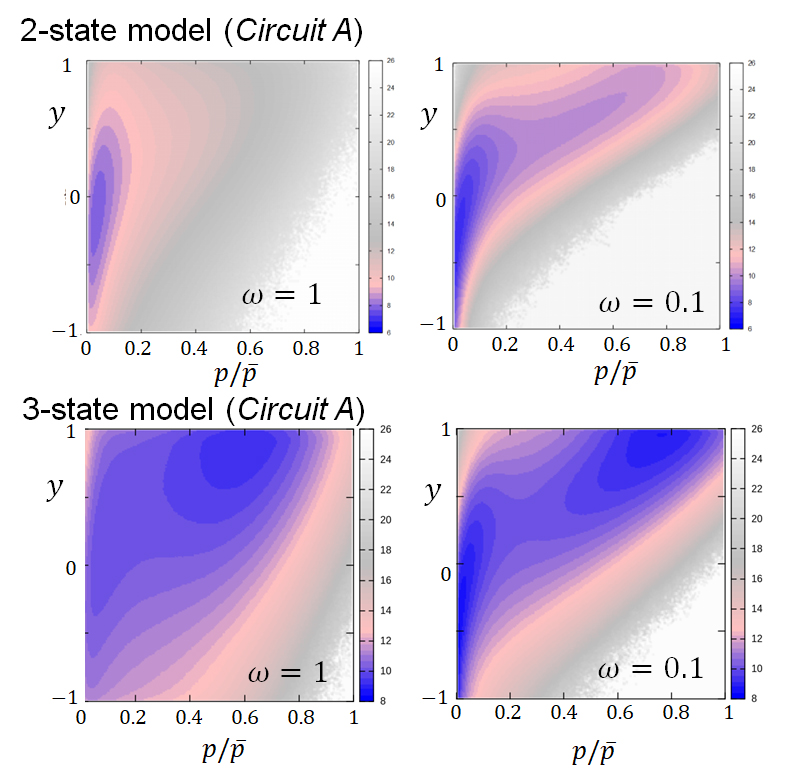}
\caption{
\changed{The landscape, $U(p,y)$, of the self-activating single-gene circuit (\textit{Circuit A}) is presented for two different models: the two-state model (Top) and three-state model (Bottom). The landscapes are compared by varying the adiabaticity $ \omega$. All other parameters remain same among panels with $\bar{p}=1$. The bottom panels are identical to those in Fig. \ref{basins1}.
}
}
\label{2stateA}
\end{figure}

\begin{figure}[htbp]
\centering
\includegraphics[width=8cm]{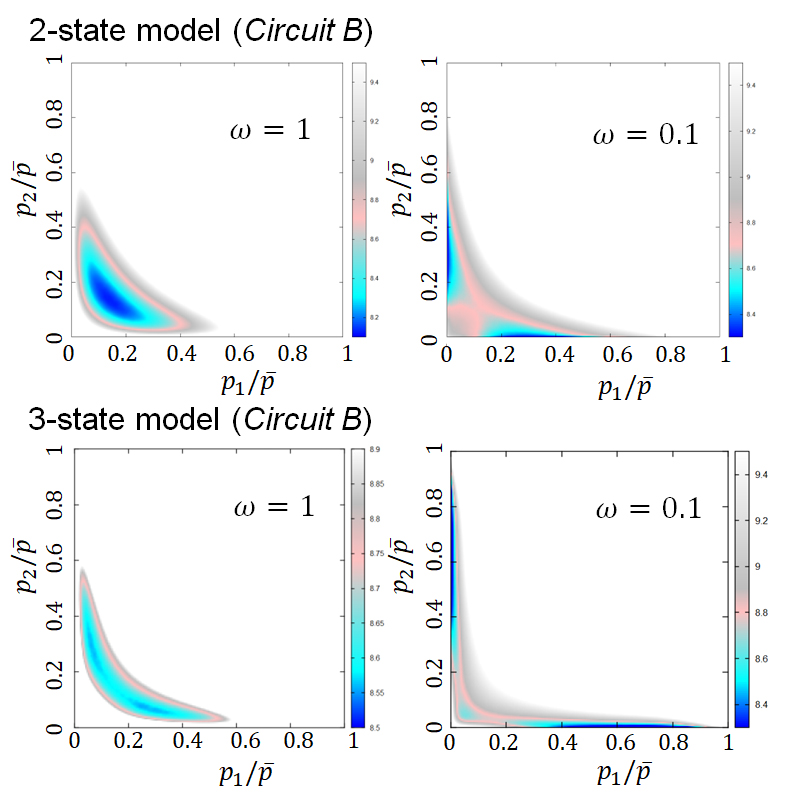}
\caption{
\changed{The landscape, $U(p_1,p_2)$, of the mutually repressing two-gene circuit (\textit{Circuit B}) is presented for two different models:  the two-state model (Top) and three-state model (Bottom).  The landscapes are compared by varying the cadiabaticity $ \omega$. All other parameters remain same among panels with $\bar{p}=1.2$. The bottom panels are identical to those in Fig. \ref{basins2}.
}
}
\label{2stateB}
\end{figure}

\changed{We refer to the original model represented by Eq.\,\ref{forceC} as the three-state model, while the model defined by Eq.\,\ref{forceC2} is referred to as the two-state model. We can straightforwardly  compare the two-state and three-state models by setting $\xi_{11}^*=\xi_{11}$, $\xi_{1-1}^*=\xi_{1-1}$, $\xi_{01}^*=\xi_{01}$, $\xi_{0-1}^*=\xi_{0-1}$, $\bar{r}^{1-1}=\bar{r}^{0-1}$, and $\bar{r}^{-11}=\bar{r}^{01}$. With the emphasis on transitions from the intermediate state ($y_i=0$) in the three-state model, both the active state ($y_i=1$) and the inactive state ($y_i=-1$) become more stabilized, leading to the coexisting multiple basins at active and inactive states. Reducing the emphasis on the intermediate state in the two-state model decreases the tendency for coexistence.
This reduced tendency can be observed in the landscapes derived from the two-state model (Figs.\,\ref{2stateA} and \ref{2stateB}).
}

\changed{
In summary, the relatively destabilized active and inactive states in the two-state model suppresses the formation of distinct multiple basins. However, both in the two-state and three-state models, reducing the adiabaticity of the chromatin-state transitions leads the system towards the emergence of multiple states.   
}

\subsection*{E: Other parameterizations for \textit{Circuit C}}

\begin{figure}[htbp]
\centering
\includegraphics[width=8cm]{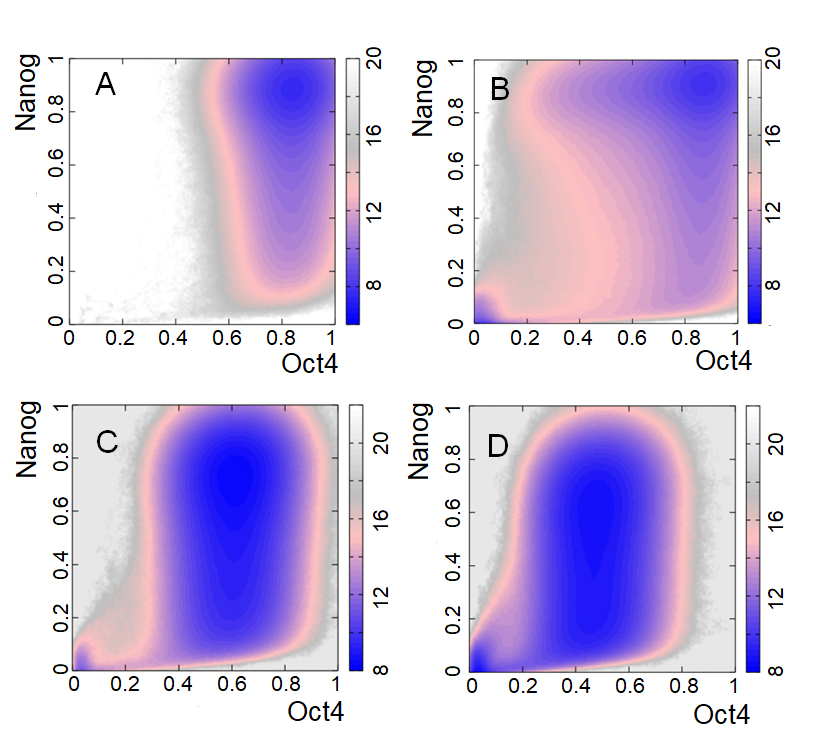}
\caption{
\changed{The landscapes of the circuit of the three core genes of mES cells, \textit{Circuit C}, calculated with various parameterizations. The two-dimensional landscapes $U(p_1, p_3)$ are plotted on the plane of  normalized concentrations of Oct4 $p_1/\bar{p}_1$ and Nanog $p_3/\bar{p}_3$. (A) The same landscape as in Fig.\,\ref{basins3hetero}. (B) The landscape calcutated with the lower adiavaticiy. (C) and (D) The landscapes calculated with the less sensitive dependence of the chromatin-state transition rates on the TF-binding state with $\gamma^{0-1}=\gamma^{10}=0.6k$ and $\gamma^{01}=\gamma^{-10}=-0.6k$ in C and $\gamma^{0-1}=\gamma^{10}=0.3k$ and $\gamma^{01}=\gamma^{-10}=-0.3k$ in D. $\gamma^{0-1}=\gamma^{10}=k$ and $\gamma^{01}=\gamma^{-10}=-k$ in A and B.
The adiabaticity is $\omega_1=\omega_2=10$ for  \textit{Oct4} and \textit{Sox2} and $\omega_3=0.5$ for \textit{Nanog} in A, C, and D and $\omega_1=\omega_2=1$ and $\omega_3=0.1$ in B. 
The typical concentrations were set to $\bar{p}_1=\bar{p}_2=\bar{p}_3=2$. 
}
}
\label{DDyna}
\end{figure}

\changed{The landscape depicted in Fig.\,\ref{basins3hetero} in Sec.\,IV  aligns with the experimental observations, demonstrating significant fluctuations in Nanog and narrow, single-peak fluctuations in Oct4. The same figure is reproduced in Fig.\,\ref{DDyna}A in the present Appendix.
However, this feature disappears under specific parameter settings as shown in Figs.\,\ref{DDyna}B--\ref{DDyna}D.
}

\changed{Fig.\,\ref{DDyna}B illustrates the landscapes generated by the lower adiabaticity for \textit{Oct4} and \textit{Sox2} (with $\omega_1 = \omega_2 = 1$). In this parameter setting, the adiabaticity of all three genes falls within the eddy regime. As a result, the landscape presents a broadened basin, similar to what is depicted in the middle panel of Fig.\,\ref{basins3}. Additionally, with the lower adiabaticity of \textit{Nanog}, an inactive state basin emerges, characterized by low protein concentrations. This inactive basin causes significant bimodal fluctuations in the concentration of \textit{Oct4}, which contradicts the experimentally observed single-peak distribution of \textit{Oct4} concentration. Consequently, distinct adiabaticity for \textit{Oct4} and \textit{Sox2}, along with nonadiabatic eddy behavior for \textit{Nanog}, is necessary to explain the experimental observations.
}

\changed{
Other examples depicted in Fig.\,\ref{DDyna}C and Fig.\,\ref{DDyna}D present cases with smaller values of $|\gamma^{yy'}|$ in Eq.\,\ref{munu}. In these scenarios, the chromatin-state transition rates show less sensitivity to the TF-binding state. Then, the feedback relations among TFs determine the landscapt to generate a basin at the inactive state. This again leads to a wide bimodal distribution in Oct4 concentration, contradicting the experimental observations.
}

\newpage
\bibliography{EuModel}

\end{document}